\documentclass[zpreprint,zbstepj]{zeus_paper}
\usepackage[english]{babel}
\usepackage{stmaryrd}

\newcommand{\ZcoosysB}{%
The ZEUS coordinate system is a right-handed Cartesian system, with the $Z$
axis pointing in the proton beam direction, referred to as the ``forward
direction'', and the $X$ axis pointing left towards the centre of HERA.
The coordinate origin is at the nominal interaction point.\xspace}
\newcommand{\Zpsrap}{%
The pseudorapidity is defined as $\eta=-\ln\left(\tan\frac{\theta}{2}\right)$,
where the polar angle, $\theta$, is measured with respect to the proton beam
direction.\xspace}

\newcommand{\Zdetdesc}{%
A detailed description of the ZEUS detector can be found 
elsewhere~\cite{zeus:1993:bluebook}. A brief outline of the 
components that are most relevant for this analysis is given
below.\xspace}

\newcommand{\Zctdmvddesc}[1]{%
In the kinematic range of the analysis, charged particles were tracked
in the central tracking detector (CTD)~\citeCTD and the microvertex
detector (MVD)~\citeMVD. These components operated in a magnetic
field of $1.43\Tesla$ provided by a thin superconducting solenoid. The
CTD consisted of 72~cylindrical drift chamber layers, organised in nine
superlayers covering the polar-angle#1 region
\mbox{$15^\circ<\theta<164^\circ$}.
The MVD silicon tracker consisted of a barrel (BMVD) and a forward
(FMVD) section. The BMVD provided polar-angle coverage for tracks with
three measurements from $30^\circ$ to $150^\circ$. The FMVD extended
the polar-angle coverage in the forward region to $7^\circ$. 
After alignment, the single-hit resolution of the BMVD was $\rm 24\,\mu m$
and the average impact parameter resolution of the CTD-BMVD system
for high-momentum tracks was $\rm 100\,\mu m$.

To estimate the ionisation energy loss per unit length, \dEdx{}, of
particles in the CTD~\cite{pl:b481:213}, the truncated mean of the
anode-wire pulse heights was calculated, which removes the lowest 10\%
and at least the highest 30\% depending on the number of saturated
hits.  The measured \dEdx{} values were corrected by normalising to
the average \dEdx{} for tracks around the region of minimum ionisation
for pions with momentum, $p$, satisfying $0.3 < p < 0.4$
GeV~\cite{thesis:bartsch:2007}.}

\newcommand{\Zsttdesc}[1]{%
The STT consisted of 48 sectors of two different sizes. Each sector
contained 192 (small sector) or 264 (large sector) straws of diameter
7.5 mm arranged into 3 layers. The sectors were trapezoidal in shape
and each subtended an azimuthal angle of $60^{\circ}$ -- 6 sectors
formed a so-called superlayer. A particle passing through the complete
detector traversed 8 superlayers, which were rotated around the beam
direction at angles of $30^{\circ}$ or $15^{\circ}$ to each other. The STT
covered the polar-angle region $5^{\circ}<\theta<23^{\circ}$.
}

\newcommand{\Zcaldesc}{%
The high-resolution uranium--scintillator calorimeter (CAL)~\citeCAL consisted
of three parts: the forward (FCAL), the barrel (BCAL) and the rear (RCAL)
calorimeters. Each part was subdivided transversely into towers and
longitudinally into one electromagnetic section and either one (in RCAL)
or two (in BCAL and FCAL) hadronic sections. The smallest subdivision of
the calorimeter is called a cell.  The CAL energy resolutions, as measured under
test-beam conditions, are $\sigma(E)/E=0.18/\sqrt{E}$ for electrons and
$\sigma(E)/E=0.35/\sqrt{E}$ for hadrons, with $E$ in $\Gev$.}

\def\citePCAL{{\cite{%
desy-92-066,*zfp:c63:391,*acpp:b32:2025%
}}\xspace}
\def\citeSPECTRO{{\cite{%
physics-0512153%
}}\xspace}
\newcommand{\Zlumidesc}{%
The luminosity was measured using the Bethe-Heitler reaction
$ep\,\rightarrow\, e\gamma p$ by a luminosity detector which consisted
of independent lead--scintillator calorimeter\citePCAL and magnetic
spectrometer\citeSPECTRO systems. 
}

\chardef\usc=95
\chardef\til=126
\catcode`\@=11 
\DeclareRobustCommand\xdotspace{\futurelet\@let@token\@xdotspace}
\def\@xdotspace{%
  \ifx\@let@token.\else
  \ifx\@let@token\bgroup.\else
  \ifx\@let@token\egroup.\else
  \ifx\@let@token\/.\else
  \ifx\@let@token\ .\else
  \ifx\@let@token~.\else
  \ifx\@let@token!.\else
  \ifx\@let@token,.\else
  \ifx\@let@token:.\else
  \ifx\@let@token;.\else
  \ifx\@let@token?.\else
  \ifx\@let@token/.\else
  \ifx\@let@token'.\else
  \ifx\@let@token).\else
  \ifx\@let@token-.\else
  \ifx\@let@token\@xobeysp.\else
  \ifx\@let@token\space.\else
  \ifx\@let@token\@sptoken.\else
   .\space
   \fi\fi\fi\fi\fi\fi\fi\fi\fi\fi\fi\fi\fi\fi\fi\fi\fi\fi}
\catcode`\@=12 

\newcommand{\stru}[2]{%
   \relax\ifmmode\hbox{\vrule height#1 depth#2 width0pt}%
   \else\vrule height#1 depth#2 width0pt\fi}

\newcommand{\Ronum}[1]{\uppercase\expandafter{\romannumeral#1}}
\newcommand{\ronum}[1]{\expandafter{\romannumeral#1}}
\DeclareRobustCommand{\LaTeXZ}{%
  \LaTeX\kern-.05em4\kern-.1em
  {\raisebox{-0.2ex}{$\scriptstyle\text{ZEUS}$}}\xspace}

\DeclareMathAlphabet{\mathbf}{OT1}{cmr}{bx}{sl}
\newcommand{\eVdist}{\kern-0.06667em}

\newcommand{\Gev}{{\text{Ge}\eVdist\text{V\/}}}

\newcommand{\gev}{{\,\text{Ge}\eVdist\text{V\/}}}

\newcommand{\pb}{\,\text{pb}}

\newcommand{\pbi}{\,\text{pb}^{-1}}

\newcommand{\cm}{\,\text{cm}}

\newcommand{\Tesla}{\,\text{T}}

\newcommand{\slashfrac}[2]{%
  \raisebox{0.5ex}{\ensuremath #1}\kern-0.12em/\kern-0.08em
  \raisebox{-.8ex}{\ensuremath #2}}

\newcommand{\sqr}[3]{%
    {\vcenter{\hrule height.#3ex\hbox{\vrule width.#2ex height#1ex
     \kern#1ex\vrule width.#3ex}\hrule height.#2ex}}}

\newcommand{\widebar}[1]{%
   \mkern1.5mu\overline{\mkern-1.5mu#1\mkern-1.mu}\mkern1.mu}
\catcode`\@=11 
\newcommand{\parenbar}{\mathpalette\p@renb@r}
\def\p@renb@r#1#2{\vbox{%
  \ifx#1\scriptscriptstyle \dimen@.7em\dimen@ii.2em\else
  \ifx#1\scriptstyle \dimen@.8em\dimen@ii.25em\else
  \dimen@1em\dimen@ii.4em\fi\fi \offinterlineskip
  \ialign{\hfill##\hfill\cr
    \vbox{\hrule width\dimen@ii}\cr
    \noalign{\vskip-.3ex}%
    \hbox to\dimen@{$\mathchar300\hfil\mathchar301$}\cr
    \noalign{\vskip-.3ex}%
    $#1#2$\cr}}}
\catcode`\@=12 

\newcommand{\cbar}{\widebar{c}}
\newcommand{\bbar}{\widebar{b}}

\newcommand{\IP}{{\rm I$\kern-0.01667em$P}\xspace}

\newcommand{\dof}{{\rm dof}}

\mathchardef\qsm=63
\mathchardef\pls=43
\mathchardef\mns=512
\mathchardef\plm=518
\mathchardef\eql=61
\mathchardef\smallleft=300
\mathchardef\smallright=301
\mathchardef\les=316
\mathchardef\gre=318
\mathchardef\leq=532
\mathchardef\grq=533

\catcode`\@=11 
\newcounter{pict@width}
\newcounter{pict@height}
\newlength{\pict@scale}
\newcommand{\psfigadd}[4]{%
\setcounter{pict@width}{1*\ratio{#2+\pict@scale/2}{\pict@scale}}
\setcounter{pict@height}{1*\ratio{#3+\pict@scale/2}{\pict@scale}}
\setlength{\unitlength}{\pict@scale}
\hbox to #2{\hspace{-\fill}\begin{picture}(\thepict@width,\thepict@height)
\put(0,0){\psfig{figure=#1,width=#2,height=#3,clip=}}
\SetScale{0.283466457}
\SetWidth{1.763889}
{#4}
\end{picture}}
}
\newcounter{pict@widthfst}
\newcounter{pict@widthscd}
\newcounter{pict@widthtot}
\newcommand{\psfigaddtwo}[7]{%
\setcounter{pict@widthfst}{1*\ratio{#2+\pict@scale/2}{\pict@scale}}
\setcounter{pict@widthscd}{1*\ratio{#2+#4+\pict@scale/2}{\pict@scale}}
\setcounter{pict@widthtot}{1*\ratio{#2+#4+#6+\pict@scale/2}{\pict@scale}}
\setcounter{pict@height}{1*\ratio{#3+\pict@scale/2}{\pict@scale}}
\setlength{\unitlength}{\pict@scale}
\hbox{\hspace{-\fill}\begin{picture}(\thepict@widthtot,\thepict@height)
\put(0,0){\psfig{figure=#1,width=#2,height=#3,clip=}}
\put(\thepict@widthscd,0){\psfig{figure=#5,width=#6,height=#3,clip=}}
\SetScale{0.283466457}
\SetWidth{1.763889}
{#7}
\end{picture}}
}
\newcommand{\psfigror}[4]{%
\setcounter{pict@width}{1*\ratio{#2+\pict@scale/2}{\pict@scale}}
\setcounter{pict@height}{1*\ratio{#3+\pict@scale/2}{\pict@scale}}
\setlength{\unitlength}{\pict@scale}
\hbox{\begin{picture}(\thepict@width,\thepict@height)
\put(0,\thepict@height){\psfig{figure=#1,width=#3,height=#2,clip=,angle=270}}
\SetScale{0.283466457}
\SetWidth{1.763889}
{#4}
\end{picture}}
}
\newcommand{\psfigrol}[4]{%
\setcounter{pict@width}{1*\ratio{#2+\pict@scale/2}{\pict@scale}}
\setcounter{pict@height}{1*\ratio{#3+\pict@scale/2}{\pict@scale}}
\setlength{\unitlength}{\pict@scale}
\hbox{\begin{picture}(\thepict@width,\thepict@height)
\put(0,0){\psfig{figure=#1,width=#3,height=#2,clip=,angle=90}}
\SetScale{0.283466457}
\SetWidth{1.763889}
{#4}
\end{picture}}
}
\catcode`\@=12 
\newlength\listtextwidth

\catcode`\@=11 
\newlength{\@tabfninsert}
\newlength{\@tabfnwidth}
\newcommand{\tabfootnote}[2]{%
  \setlength{\@tabfninsert}{0.8em}
  \setlength{\@tabfnwidth}{\textwidth}
  \addtolength{\@tabfnwidth}{-\@tabfninsert}
  \addtolength{\@tabfnwidth}{-0.4em}
  \noindent\makebox[\@tabfninsert][r]{\footnotesize$^{#1}$\hfil}\hfill%
  \parbox[t]{\@tabfnwidth}{\footnotesize #2\hfill}}
\catcode`\@=12 

\def\citeCTD{{\cite{%
nim:a279:290,*npps:b32:181,*nim:a338:254%
}}\xspace}
\def\citeCAL{{\cite{%
nim:a309:77,*nim:a309:101,*nim:a321:356,*nim:a336:23%
}}\xspace}
\def\citeMVD{{\cite{%
nim:a581:656%
}}\xspace}

\newcommand{\ZcoosysfnBetaphi}{\footnote{\ZcoosysB\Zpsrap{} The
    azimuthal angle, $\phi$, is measured with respect to the $X$ axis.}}

\newcommand{\Qsq}{\ensuremath{{Q^{2}}}}

\newcommand{\etajet}{\ensuremath{\eta^{\textrm{jet}}}}

\newcommand{\phijet}{\ensuremath{\phi^{\textrm{jet}}}}
\newcommand{\pTrel}{\ensuremath{p_{T}^{\textrm{rel}}}}
\newcommand{\DLsig}{\ensuremath{d/\delta d}}

\newcommand{\pTe}{\ensuremath{p_{T}^{e}}}
\newcommand{\etae}{\ensuremath{\eta^{e}}}
\newcommand{\pTmiss}{\ensuremath{\vec{\not\!p}_{T}}}
\newcommand{\Dphi}{\ensuremath{\Delta\phi}}

\newcommand{\btoe}{\ensuremath{b \shortrightarrow e}}
\newcommand{\othe}{\ensuremath{\text{other }e}}
\newcommand{\ebkg}{\ensuremath{\text{non-}e}}

\newcommand{\Jpsi}{\ensuremath{J/\psi}}

\newcommand{\signlob}{\ensuremath{\sigma^{\textrm{NLO}}}}

\newcommand{\dEdx}{\ensuremath{\mathrm{d}E/\mathrm{d}x}}

\newcommand{\dif}{\ensuremath{\mathrm{d}}}
\newcommand{\stat}{\ensuremath{\text{stat.}}}
\newcommand{\syst}{\ensuremath{\text{syst.}}}
\newcommand{\PYTHIA}{\textsc{Pythia}}
\newcommand{\RAPGAP}{\textsc{Rapgap}}
\newcommand{\DJANGOH}{\textsc{Djangoh}}
\newcommand{\ARIADNE}{\textsc{Ariadne}}
\newcommand{\HERACLES}{\textsc{Heracles}}
\newcommand{\JETSET}{\textsc{Jetset}}
\begin{document}
\prepnum{DESY--11--005}
\title{Measurement of beauty production in deep inelastic scattering at HERA using decays into electrons}
                    
\author{ZEUS Collaboration}
\date{19 January 2011}

\abstract{The production of beauty quarks in $ep$ interactions has
  been studied with the ZEUS detector at HERA for exchanged
  four-momentum squared $Q^{2} > 10\gev^{2}$, using an
  integrated luminosity of $363\pbi$. The beauty events were
  identified using electrons from semileptonic $b$ decays with a
  transverse momentum $0.9 < p_{T}^{e} < 8\gev$ and
  pseudorapidity $|\eta^{e}|<1.5.$ Cross sections for beauty
  production were measured and compared with next-to-leading-order QCD
  calculations. The beauty contribution to the
  proton structure function $F_{2}$ was extracted from
  the double-differential cross section as a function of Bjorken-$x$
  and $Q^{2}$.}

\makezeustitle
\def\3{\ss}
\pagenumbering{Roman}

\begin{center}
{ \Large The ZEUS Collaboration }
\end{center}

{\small


{\mbox H.~Abramowicz$^{45, ah}$, }
{\mbox I.~Abt$^{35}$, }
{\mbox L.~Adamczyk$^{13}$, }
{\mbox M.~Adamus$^{54}$, }
{\mbox R.~Aggarwal$^{7, d}$, }
{\mbox S.~Antonelli$^{4}$, }
{\mbox P.~Antonioli$^{3}$, }
{\mbox A.~Antonov$^{33}$, }
{\mbox M.~Arneodo$^{50}$, }
{\mbox V.~Aushev$^{26, 27, aa}$
% ...
}
\newpage


\makebox[3em]{$^{1}$}
\begin{minipage}[t]{14cm}
{\it Argonne National Laboratory, Argonne, Illinois 60439-4815, USA}~$^{A}$

\end{minipage}\\
\makebox[3em]{$^{2}$}
\begin{minipage}[t]{14cm}
{\it Andrews University, Berrien Springs, Michigan 49104-0380, USA}

\end{minipage}\\
\makebox[3em]{$^{3}$}
\begin{minipage}[t]{14cm}
{\it INFN Bologna, Bologna, Italy}~$^{B}$

\end{minipage}\\
\makebox[3em]{$^{4}$}
\begin{minipage}[t]{14cm}
{\it University and INFN Bologna, Bologna, Italy}~$^{B}$

\end{minipage}\\
\makebox[3em]{$^{5}$}
\begin{minipage}[t]{14cm}
{\it Physikalisches Institut der Universit\"at Bonn,
Bonn, Germany}~$^{C}$

\end{minipage}\\
\makebox[3em]{$^{6}$}
\begin{minipage}[t]{14cm}
{\it H.H.~Wills Physics Laboratory, University of Bristol,
Bristol, United Kingdom}~$^{D}$

\end{minipage}\\
\makebox[3em]{$^{7}$}
\begin{minipage}[t]{14cm}
{\it Panjab University, Department of Physics, Chandigarh, India}

\end{minipage}\\
\makebox[3em]{$^{8}$}
\begin{minipage}[t]{14cm}
{\it Calabria University,
Physics Department and INFN, Cosenza, Italy}~$^{B}$

\end{minipage}\\
\makebox[3em]{$^{9}$}
\begin{minipage}[t]{14cm}
{\it Institute for Universe and Elementary Particles, Chonnam National University,\\
Kwangju, South Korea}

\end{minipage}\\
\makebox[3em]{$^{10}$}
\begin{minipage}[t]{14cm}
{\it Jabatan Fizik, Universiti Malaya, 50603 Kuala Lumpur, Malaysia}~$^{E}$

\end{minipage}\\
\makebox[3em]{$^{11}$}
\begin{minipage}[t]{14cm}
{\it Nevis Laboratories, Columbia University, Irvington on Hudson,
New York 10027, USA}~$^{F}$

\end{minipage}\\
\makebox[3em]{$^{12}$}
\begin{minipage}[t]{14cm}
{\it The Henryk Niewodniczanski Institute of Nuclear Physics, Polish Academy of \\
Sciences, Cracow, Poland}~$^{G}$

\end{minipage}\\
\makebox[3em]{$^{13}$}
\begin{minipage}[t]{14cm}
{\it Faculty of Physics and Applied Computer Science, AGH-University of Science and \\
Technology, Cracow, Poland}~$^{H}$

\end{minipage}\\
\makebox[3em]{$^{14}$}
\begin{minipage}[t]{14cm}
{\it Department of Physics, Jagellonian University, Cracow, Poland}

\end{minipage}\\
\makebox[3em]{$^{15}$}
\begin{minipage}[t]{14cm}
{\it Deutsches Elektronen-Synchrotron DESY, Hamburg, Germany}

\end{minipage}\\
\makebox[3em]{$^{16}$}
\begin{minipage}[t]{14cm}
{\it Deutsches Elektronen-Synchrotron DESY, Zeuthen, Germany}

\end{minipage}\\
\makebox[3em]{$^{17}$}
\begin{minipage}[t]{14cm}
{\it INFN Florence, Florence, Italy}~$^{B}$

\end{minipage}\\
\makebox[3em]{$^{18}$}
\begin{minipage}[t]{14cm}
{\it University and INFN Florence, Florence, Italy}~$^{B}$

\end{minipage}\\
\makebox[3em]{$^{19}$}
\begin{minipage}[t]{14cm}
{\it Fakult\"at f\"ur Physik der Universit\"at Freiburg i.Br.,
Freiburg i.Br., Germany}

\end{minipage}\\
\makebox[3em]{$^{20}$}
\begin{minipage}[t]{14cm}
{\it School of Physics and Astronomy, University of Glasgow,
Glasgow, United Kingdom}~$^{D}$

\end{minipage}\\
\makebox[3em]{$^{21}$}
\begin{minipage}[t]{14cm}
{\it Department of Engineering in Management and Finance, Univ. of
the Aegean, Chios, Greece}

\end{minipage}\\
\makebox[3em]{$^{22}$}
\begin{minipage}[t]{14cm}
{\it Hamburg University, Institute of Experimental Physics, Hamburg,
Germany}~$^{I}$

\end{minipage}\\
\makebox[3em]{$^{23}$}
\begin{minipage}[t]{14cm}
{\it Imperial College London, High Energy Nuclear Physics Group,
London, United Kingdom}~$^{D}$

\end{minipage}\\
\makebox[3em]{$^{24}$}
\begin{minipage}[t]{14cm}
{\it Institute of Particle and Nuclear Studies, KEK,
Tsukuba, Japan}~$^{J}$

\end{minipage}\\
\makebox[3em]{$^{25}$}
\begin{minipage}[t]{14cm}
{\it Institute of Physics and Technology of Ministry of Education and
Science of Kazakhstan, Almaty, Kazakhstan}

\end{minipage}\\
\makebox[3em]{$^{26}$}
\begin{minipage}[t]{14cm}
{\it Institute for Nuclear Research, National Academy of Sciences, Kyiv, Ukraine}

\end{minipage}\\
\makebox[3em]{$^{27}$}
\begin{minipage}[t]{14cm}
{\it Department of Nuclear Physics, National Taras Shevchenko University of Kyiv, Kyiv, Ukraine}

\end{minipage}\\
\makebox[3em]{$^{28}$}
\begin{minipage}[t]{14cm}
{\it Kyungpook National University, Center for High Energy Physics, Daegu,
South Korea}~$^{K}$

\end{minipage}\\
\makebox[3em]{$^{29}$}
\begin{minipage}[t]{14cm}
{\it Institut de Physique Nucl\'{e}aire, Universit\'{e} Catholique de Louvain, Louvain-la-Neuve,\\
Belgium}~$^{L}$

\end{minipage}\\
\makebox[3em]{$^{30}$}
\begin{minipage}[t]{14cm}
{\it Departamento de F\'{\i}sica Te\'orica, Universidad Aut\'onoma
de Madrid, Madrid, Spain}~$^{M}$

\end{minipage}\\
\makebox[3em]{$^{31}$}
\begin{minipage}[t]{14cm}
{\it Department of Physics, McGill University,
Montr\'eal, Qu\'ebec, Canada H3A 2T8}~$^{N}$

\end{minipage}\\
\makebox[3em]{$^{32}$}
\begin{minipage}[t]{14cm}
{\it Meiji Gakuin University, Faculty of General Education,
Yokohama, Japan}~$^{J}$

\end{minipage}\\
\makebox[3em]{$^{33}$}
\begin{minipage}[t]{14cm}
{\it Moscow Engineering Physics Institute, Moscow, Russia}~$^{O}$

\end{minipage}\\
\makebox[3em]{$^{34}$}
\begin{minipage}[t]{14cm}
{\it Moscow State University, Institute of Nuclear Physics,
Moscow, Russia}~$^{P}$

\end{minipage}\\
\makebox[3em]{$^{35}$}
\begin{minipage}[t]{14cm}
{\it Max-Planck-Institut f\"ur Physik, M\"unchen, Germany}

\end{minipage}\\
\makebox[3em]{$^{36}$}
\begin{minipage}[t]{14cm}
{\it NIKHEF and University of Amsterdam, Amsterdam, Netherlands}~$^{Q}$

\end{minipage}\\
\makebox[3em]{$^{37}$}
\begin{minipage}[t]{14cm}
{\it Physics Department, Ohio State University,
Columbus, Ohio 43210, USA}~$^{A}$

\end{minipage}\\
\makebox[3em]{$^{38}$}
\begin{minipage}[t]{14cm}
{\it Department of Physics, University of Oxford,
Oxford, United Kingdom}~$^{D}$

\end{minipage}\\
\makebox[3em]{$^{39}$}
\begin{minipage}[t]{14cm}
{\it INFN Padova, Padova, Italy}~$^{B}$

\end{minipage}\\
\makebox[3em]{$^{40}$}
\begin{minipage}[t]{14cm}
{\it Dipartimento di Fisica dell' Universit\`a and INFN,
Padova, Italy}~$^{B}$

\end{minipage}\\
\makebox[3em]{$^{41}$}
\begin{minipage}[t]{14cm}
{\it Department of Physics, Pennsylvania State University, University Park,\\
Pennsylvania 16802, USA}~$^{F}$

\end{minipage}\\
\makebox[3em]{$^{42}$}
\begin{minipage}[t]{14cm}
{\it Polytechnic University, Sagamihara, Japan}~$^{J}$

\end{minipage}\\
\makebox[3em]{$^{43}$}
\begin{minipage}[t]{14cm}
{\it Dipartimento di Fisica, Universit\`a 'La Sapienza' and INFN,
Rome, Italy}~$^{B}$

\end{minipage}\\
\makebox[3em]{$^{44}$}
\begin{minipage}[t]{14cm}
{\it Rutherford Appleton Laboratory, Chilton, Didcot, Oxon,
United Kingdom}~$^{D}$

\end{minipage}\\
\makebox[3em]{$^{45}$}
\begin{minipage}[t]{14cm}
{\it Raymond and Beverly Sackler Faculty of Exact Sciences, School of Physics, \\
Tel Aviv University, Tel Aviv, Israel}~$^{R}$

\end{minipage}\\
\makebox[3em]{$^{46}$}
\begin{minipage}[t]{14cm}
{\it Department of Physics, Tokyo Institute of Technology,
Tokyo, Japan}~$^{J}$

\end{minipage}\\
\makebox[3em]{$^{47}$}
\begin{minipage}[t]{14cm}
{\it Department of Physics, University of Tokyo,
Tokyo, Japan}~$^{J}$

\end{minipage}\\
\makebox[3em]{$^{48}$}
\begin{minipage}[t]{14cm}
{\it Tokyo Metropolitan University, Department of Physics,
Tokyo, Japan}~$^{J}$

\end{minipage}\\
\makebox[3em]{$^{49}$}
\begin{minipage}[t]{14cm}
{\it Universit\`a di Torino and INFN, Torino, Italy}~$^{B}$

\end{minipage}\\
\makebox[3em]{$^{50}$}
\begin{minipage}[t]{14cm}
{\it Universit\`a del Piemonte Orientale, Novara, and INFN, Torino,
Italy}~$^{B}$

\end{minipage}\\
\makebox[3em]{$^{51}$}
\begin{minipage}[t]{14cm}
{\it Department of Physics, University of Toronto, Toronto, Ontario,
Canada M5S 1A7}~$^{N}$

\end{minipage}\\
\makebox[3em]{$^{52}$}
\begin{minipage}[t]{14cm}
{\it Physics and Astronomy Department, University College London,
London, United Kingdom}~$^{D}$

\end{minipage}\\
\makebox[3em]{$^{53}$}
\begin{minipage}[t]{14cm}
{\it Faculty of Physics, University of Warsaw, Warsaw, Poland}

\end{minipage}\\
\makebox[3em]{$^{54}$}
\begin{minipage}[t]{14cm}
{\it Institute for Nuclear Studies, Warsaw, Poland}

\end{minipage}\\
\makebox[3em]{$^{55}$}
\begin{minipage}[t]{14cm}
{\it Department of Particle Physics and Astrophysics, Weizmann
Institute, Rehovot, Israel}

\end{minipage}\\
\makebox[3em]{$^{56}$}
\begin{minipage}[t]{14cm}
{\it Department of Physics, University of Wisconsin, Madison,
Wisconsin 53706, USA}~$^{A}$

\end{minipage}\\
\makebox[3em]{$^{57}$}
\begin{minipage}[t]{14cm}
{\it Department of Physics, York University, Ontario, Canada M3J
1P3}~$^{N}$

\end{minipage}\\
\vspace{30em} \pagebreak[4]


\makebox[3ex]{$^{ A}$}
\begin{minipage}[t]{14cm}
 supported by the US Department of Energy\
\end{minipage}\\
\makebox[3ex]{$^{ B}$}
\begin{minipage}[t]{14cm}
 supported by the Italian National Institute for Nuclear Physics (INFN) \
\end{minipage}\\
\makebox[3ex]{$^{ C}$}
\begin{minipage}[t]{14cm}
 supported by the German Federal Ministry for Education and Research (BMBF), under
 contract No. 05 H09PDF\
\end{minipage}\\
\makebox[3ex]{$^{ D}$}
\begin{minipage}[t]{14cm}
 supported by the Science and Technology Facilities Council, UK\
\end{minipage}\\
\makebox[3ex]{$^{ E}$}
\begin{minipage}[t]{14cm}
 supported by an FRGS grant from the Malaysian government\
\end{minipage}\\
\makebox[3ex]{$^{ F}$}
\begin{minipage}[t]{14cm}
 supported by the US National Science Foundation. Any opinion,
 findings and conclusions or recommendations expressed in this material
 are those of the authors and do not necessarily reflect the views of the
 National Science Foundation.\
\end{minipage}\\
\makebox[3ex]{$^{ G}$}
\begin{minipage}[t]{14cm}
 supported by the Polish Ministry of Science and Higher Education as a scientific project No.
 DPN/N188/DESY/2009\
\end{minipage}\\
\makebox[3ex]{$^{ H}$}
\begin{minipage}[t]{14cm}
 supported by the Polish Ministry of Science and Higher Education
 as a scientific project (2009-2010)\
\end{minipage}\\
\makebox[3ex]{$^{ I}$}
\begin{minipage}[t]{14cm}
 supported by the German Federal Ministry for Education and Research (BMBF), under
 contract No. 05h09GUF, and the SFB 676 of the Deutsche Forschungsgemeinschaft (DFG) \
\end{minipage}\\
\makebox[3ex]{$^{ J}$}
\begin{minipage}[t]{14cm}
 supported by the Japanese Ministry of Education, Culture, Sports, Science and Technology
 (MEXT) and its grants for Scientific Research\
\end{minipage}\\
\makebox[3ex]{$^{ K}$}
\begin{minipage}[t]{14cm}
 supported by the Korean Ministry of Education and Korea Science and Engineering
 Foundation\
\end{minipage}\\
\makebox[3ex]{$^{ L}$}
\begin{minipage}[t]{14cm}
 supported by FNRS and its associated funds (IISN and FRIA) and by an Inter-University
 Attraction Poles Programme subsidised by the Belgian Federal Science Policy Office\
\end{minipage}\\
\makebox[3ex]{$^{ M}$}
\begin{minipage}[t]{14cm}
 supported by the Spanish Ministry of Education and Science through funds provided by
 CICYT\
\end{minipage}\\
\makebox[3ex]{$^{ N}$}
\begin{minipage}[t]{14cm}
 supported by the Natural Sciences and Engineering Research Council of Canada (NSERC) \
\end{minipage}\\
\makebox[3ex]{$^{ O}$}
\begin{minipage}[t]{14cm}
 partially supported by the German Federal Ministry for Education and Research (BMBF)\
\end{minipage}\\
\makebox[3ex]{$^{ P}$}
\begin{minipage}[t]{14cm}
 supported by RF Presidential grant N 41-42.2010.2 for the Leading
 Scientific Schools and by the Russian Ministry of Education and Science through its
 grant for Scientific Research on High Energy Physics\
\end{minipage}\\
\makebox[3ex]{$^{ Q}$}
\begin{minipage}[t]{14cm}
 supported by the Netherlands Foundation for Research on Matter (FOM)\
\end{minipage}\\
\makebox[3ex]{$^{ R}$}
\begin{minipage}[t]{14cm}
 supported by the Israel Science Foundation\
\end{minipage}\\
\vspace{30em} \pagebreak[4]


\makebox[3ex]{$^{ a}$}
\begin{minipage}[t]{14cm}
also affiliated with University College London,
 United Kingdom\
\end{minipage}\\
\makebox[3ex]{$^{ b}$}
\begin{minipage}[t]{14cm}
now at University of Salerno, Italy\
\end{minipage}\\
\makebox[3ex]{$^{ c}$}
\begin{minipage}[t]{14cm}
now at Queen Mary University of London, United Kingdom\
\end{minipage}\\
\makebox[3ex]{$^{ d}$}
\begin{minipage}[t]{14cm}
also funded by Max Planck Institute for Physics, Munich, Germany\
\end{minipage}\\
\makebox[3ex]{$^{ e}$}
\begin{minipage}[t]{14cm}
also Senior Alexander von Humboldt Research Fellow at Hamburg University,
 Institute of Experimental Physics, Hamburg, Germany\
\end{minipage}\\
\makebox[3ex]{$^{ f}$}
\begin{minipage}[t]{14cm}
also at Cracow University of Technology, Faculty of Physics,
 Mathemathics and Applied Computer Science, Poland\
\end{minipage}\\
\makebox[3ex]{$^{ g}$}
\begin{minipage}[t]{14cm}
supported by the research grant No. 1 P03B 04529 (2005-2008)\
\end{minipage}\\
\makebox[3ex]{$^{ h}$}
\begin{minipage}[t]{14cm}
now at Rockefeller University, New York, NY
 10065, USA\
\end{minipage}\\
\makebox[3ex]{$^{ i}$}
\begin{minipage}[t]{14cm}
now at DESY group FS-CFEL-1\
\end{minipage}\\
\makebox[3ex]{$^{ j}$}
\begin{minipage}[t]{14cm}
now at Institute of High Energy Physics, Beijing, China\
\end{minipage}\\
\makebox[3ex]{$^{ k}$}
\begin{minipage}[t]{14cm}
now at DESY group FEB, Hamburg, Germany\
\end{minipage}\\
\makebox[3ex]{$^{ l}$}
\begin{minipage}[t]{14cm}
also at Moscow State University, Russia\
\end{minipage}\\
\makebox[3ex]{$^{ m}$}
\begin{minipage}[t]{14cm}
now at University of Liverpool, United Kingdom\
\end{minipage}\\
\makebox[3ex]{$^{ n}$}
\begin{minipage}[t]{14cm}
now at CERN, Geneva, Switzerland\
\end{minipage}\\
\makebox[3ex]{$^{ o}$}
\begin{minipage}[t]{14cm}
also affiliated with Universtiy College London, UK\
\end{minipage}\\
\makebox[3ex]{$^{ p}$}
\begin{minipage}[t]{14cm}
now at Goldman Sachs, London, UK\
\end{minipage}\\
\makebox[3ex]{$^{ q}$}
\begin{minipage}[t]{14cm}
also at Institute of Theoretical and Experimental Physics, Moscow, Russia\
\end{minipage}\\
\makebox[3ex]{$^{ r}$}
\begin{minipage}[t]{14cm}
also at INP, Cracow, Poland\
\end{minipage}\\
\makebox[3ex]{$^{ s}$}
\begin{minipage}[t]{14cm}
also at FPACS, AGH-UST, Cracow, Poland\
\end{minipage}\\
\makebox[3ex]{$^{ t}$}
\begin{minipage}[t]{14cm}
partially supported by Warsaw University, Poland\
\end{minipage}\\
\makebox[3ex]{$^{ u}$}
\begin{minipage}[t]{14cm}
now at Istituto Nucleare di Fisica Nazionale (INFN), Pisa, Italy\
\end{minipage}\\
\makebox[3ex]{$^{ v}$}
\begin{minipage}[t]{14cm}
now at Haase Energie Technik AG, Neum\"unster, Germany\
\end{minipage}\\
\makebox[3ex]{$^{ w}$}
\begin{minipage}[t]{14cm}
now at Department of Physics, University of Bonn, Germany\
\end{minipage}\\
\makebox[3ex]{$^{ x}$}
\begin{minipage}[t]{14cm}
also affiliated with DESY, Germany\
\end{minipage}\\
\makebox[3ex]{$^{ y}$}
\begin{minipage}[t]{14cm}
also at University of Tokyo, Japan\
\end{minipage}\\
\makebox[3ex]{$^{ z}$}
\begin{minipage}[t]{14cm}
now at Kobe University, Japan\
\end{minipage}\\
\makebox[3ex]{$^{\dagger}$}
\begin{minipage}[t]{14cm}
 deceased \
\end{minipage}\\
\makebox[3ex]{$^{aa}$}
\begin{minipage}[t]{14cm}
supported by DESY, Germany\
\end{minipage}\\
\makebox[3ex]{$^{ab}$}
\begin{minipage}[t]{14cm}
member of National Technical University of Ukraine, Kyiv Polytechnic Institute, Kyiv,
 Ukraine\
\end{minipage}\\
\makebox[3ex]{$^{ac}$}
\begin{minipage}[t]{14cm}
member of National University of Kyiv - Mohyla Academy, Kyiv, Ukraine\
\end{minipage}\\
\makebox[3ex]{$^{ad}$}
\begin{minipage}[t]{14cm}
supported by the Bogolyubov Institute for Theoretical Physics of the National
 Academy of Sciences, Ukraine\
\end{minipage}\\
\makebox[3ex]{$^{ae}$}
\begin{minipage}[t]{14cm}
STFC Advanced Fellow\
\end{minipage}\\
\makebox[3ex]{$^{af}$}
\begin{minipage}[t]{14cm}
nee Korcsak-Gorzo\
\end{minipage}\\
\makebox[3ex]{$^{ag}$}
\begin{minipage}[t]{14cm}
This material was based on work supported by the
 National Science Foundation, while working at the Foundation.\
\end{minipage}\\
\makebox[3ex]{$^{ah}$}
\begin{minipage}[t]{14cm}
also at Max Planck Institute for Physics, Munich, Germany, External Scientific Member\
\end{minipage}\\
\makebox[3ex]{$^{ai}$}
\begin{minipage}[t]{14cm}
now at Tokyo Metropolitan University, Japan\
\end{minipage}\\
\makebox[3ex]{$^{aj}$}
\begin{minipage}[t]{14cm}
now at Nihon Institute of Medical Science, Japan\
\end{minipage}\\
\makebox[3ex]{$^{ak}$}
\begin{minipage}[t]{14cm}
now at Osaka University, Osaka, Japan\
\end{minipage}\\
\makebox[3ex]{$^{al}$}
\begin{minipage}[t]{14cm}
also at \L\'{o}d\'{z} University, Poland\
\end{minipage}\\
\makebox[3ex]{$^{am}$}
\begin{minipage}[t]{14cm}
member of \L\'{o}d\'{z} University, Poland\
\end{minipage}\\
\makebox[3ex]{$^{an}$}
\begin{minipage}[t]{14cm}
now at Lund University, Lund, Sweden\
\end{minipage}\\
\makebox[3ex]{$^{ao}$}
\begin{minipage}[t]{14cm}
also at University of Podlasie, Siedlce, Poland\
\end{minipage}\\

}

\clearpage

\pagenumbering{arabic} 
\pagestyle{plain}
\section{Introduction}
\label{sec:int}

The production of heavy quarks in $ep$ collisions at HERA is an
important testing ground for perturbative Quantum Chromodynamics
(pQCD), since the large $b$-quark mass provides a hard scale that
allows perturbative calculations to be made~\cite{np:b392:162,
np:b392:229}.  The dominant production process is boson-gluon fusion
(BGF) between the incoming virtual photon and a gluon in the proton.
Beauty production has been measured using several methods by the
ZEUS~\cite{Abramowicz:2010zq,Chekanov:2009kj,Chekanov:2008tx,Chekanov:2008zz,Chekanov:2008aaa,epj:c50:1434,pl:b599:173,pr:d70:012008,epj:c18:625}
and the
H1~\cite{:2009ut,epj:c47:597,epj:c45:23,pl:b621:56,epj:c40:349,epj:c41:453,pl:b467:156}
collaborations both in deep inelastic scattering (DIS), i.e.\ for
large exchanged four-momentum squared, $Q^{2}$, and also in
photoproduction, i.e.\ for $Q^{2} \sim 0\gev^2$.  The measurements are
reasonably well described by next-to-leading-order (NLO) QCD
predictions.

Most of the previous measurements of $b$-quark production used muons to tag
semileptonic decays of the $B$ hadrons.
This paper reports a measurement of beauty production in DIS using
the semileptonic decays to electrons,
\begin{equation*}
ep \rightarrow e'\, b \bbar \,X \rightarrow e'\, e \, X',
\end{equation*}
in the kinematic range $\Qsq>10\gev^2$. Using the electron channel allows
a measurement of the decay leptons at lower transverse momentum and provides
a complementary measurement, with independent systematics.

An analysis of the same process in the photoproduction regime, based
on data taken in 1996--2000 ($120\pbi$), used a likelihood-ratio test
to extract the signal of beauty and charm semileptonic decays to
electrons~\cite{Chekanov:2008aaa}. A similar method, adapted to the
different kinematics of the DIS regime, was used for the measurement
reported here. The analysis also benefited from improved tracking
in the more recent data, which allowed the measured decay length of
weakly decaying $B$ hadrons to be used.

In this analysis, the total visible cross section, $\sigma_{b
  \shortrightarrow e}$, and differential cross sections as a function
of $Q^{2}$, the Bjorken scaling variable, $x$, the transverse
momentum, $p_{T}^{e}$, and the pseudorapidity of the electron,
$\eta^{e}$, were measured. They are compared to a leading-order (LO)
plus parton-shower (PS) Monte Carlo prediction and to an NLO QCD
calculation. The beauty contribution to the proton structure function $F_{2}$,
denoted as $F_{2}^{b\bar{b}}$, was extracted from the double-differential cross
section as a function of $Q^{2}$ and $x$ and is compared with
theoretical calculations.

\section{Experimental set-up}
\label{sec:exp}

This analysis was performed with data taken from 2004 to 2007, when
HERA collided electrons or positrons with energy $E_e=27.5\gev$ with
protons of an energy of 920\gev, corresponding to a centre-of-mass
energy $\sqrt{s} = 318\gev$. This data-taking period is denoted as HERA~II.
The corresponding integrated luminosity is $(363 \pm 7)\pbi$.

\Zdetdesc 

\Zctdmvddesc\ZcoosysfnBetaphi\   

\Zcaldesc

\Zlumidesc

\section{Monte Carlo simulation}
\label{sec:montecarlo}

To evaluate the detector acceptance and to provide the signal and
background distributions for the likelihood-ratio test, Monte Carlo
(MC) samples of beauty, charm and light-flavour events were generated,
corresponding to eighteen, two and one times the integrated luminosity
of the data, respectively.  The \RAPGAP{} 3.00 Monte Carlo
program~\cite{cpc:86:147} was used to generate the beauty and charm
samples.  The CTEQ5L~\cite{epj:c12:375} parton density functions were
used and the heavy-quark masses were set to $m_{b}=4.75\gev$ and
$m_{c}=1.5\gev$. To simulate radiative corrections, the events were
passed through the \HERACLES{} 4.6~\cite{cpc:69:155} program.  An
inclusive MC sample containing all flavours was generated using
\DJANGOH{} 1.6~\cite{proc:hera:1991:1419} interfaced to \ARIADNE{}
4.12~\cite{cpc:71:15}, where the quarks were taken to be massless. The
CTEQ5D~\cite{epj:c12:375} parton density functions were used.

For the acceptance determination, the $Q^{2}$ distribution  
in the signal MC was reweighted in order to correct for observed
differences between the measured and simulated distributions. The corrections
varied from +10\% at low $Q^{2}$ to $-$30\% at high $Q^{2}$.
The $B$-hadron lifetimes were corrected for
differences between the simulated values and the world-average
values~\cite{physlett:b667:1}.

Fragmentation and particle decays were simulated using the
\JETSET{}/\PYTHIA{} model~\cite{cpc:82:74,*cpc:135:238}. The lepton
energy spectrum from charm decays was reweighted to agree with CLEO
data~\cite{prl:97:251801}.  The generated events were passed through a
full simulation of the ZEUS detector based on \textsc{Geant
  3.21}~\cite{tech:cern-dd-ee-84-1}.  The final MC events had
to fulfil the same trigger requirements and pass the same
reconstruction program as the data.

\section{Theoretical predictions and uncertainties}
\label{sec:theory}

Next-to-leading-order QCD predictions were obtained from the
HVQDIS~\cite{pr:d57:2806} program in the fixed-flavour-number scheme
(FFNS)\cite{np:b374:36}. More details about
the calculation can be found elsewhere~\cite{Chekanov:2009kj}.

The $b$-quark mass (pole mass) was set to $m_{b}=4.75\gev$.  The
renormalisation and factorisation scales, $\mu_{R}$ and $\mu_{F}$,
were chosen to be equal and set to
$\mu_{R}=\mu_{F}=\sqrt{Q^{2}+4m_{b}^{2}}$. The parton density
functions were obtained from the FFNS variant of the ZEUS-S
fit~\cite{pr:d67:012007} using the same $b$-quark mass as in the
HVQDIS calculation. The value of $\alpha_{s}(M_{Z})$ was set to
0.105.

The Peterson fragmentation function~\cite{pr:d27:105}, with
$\epsilon_{b}=0.0035$~\cite{Nason2000245}, was used to produce beauty
hadrons from the heavy quarks. 
The semileptonic decay spectrum was taken from the \PYTHIA{} Monte Carlo.
The contributions from prompt and from cascade decays, $b \rightarrow c (\cbar)
\rightarrow e$, including $b \rightarrow \tau
\rightarrow e$ and $b \rightarrow \Jpsi \rightarrow e^{+}e^{-}$, were
taken into account in the effective branching fraction, which was set
to 0.217\cite{physlett:b667:1}.

To estimate the uncertainty on the theoretical predictions, the
$b$-quark mass was varied in the range $m_{b} = 4.5,5.0\gev$, and the
scales $\mu_{R}$, $\mu_{F}$ were varied independently by a factor of two
up and down.  The parameter $\epsilon_{b}$ was varied by $\pm 0.002$.
The parton density functions were varied within the total
uncertainties of the fit.  The uncertainty on the NLO QCD prediction
for the total cross section is $+15\%$ and $-16\%$, where the
dominant contribution originates from the variation
of the mass and the scales.

The HVQDIS calculations were also used to extrapolate the visible
cross sections to $F_{2}^{b\bar{b}}$.

\section{Data selection}
\label{sec:data}

Events were selected online with a three-level
trigger~\cite{zeus:1993:bluebook,proc:chep:1992:222} using a
combination of triggers, which required a scattered electron to be
detected in the CAL and/or the presence of an electron candidate from
a semileptonic decay.  Further details on the trigger chain can be
found elsewhere~\cite{thesis:shehzadi:2011}.  Offline, the
reconstructed scattered electron was required to have an energy
$E_{e'}>10\gev$. The $Z$ position of the primary vertex had to be
within $|Z_{\mathrm{vtx}}| < 30\cm$.

The final state of the electron--proton collision, including the
scattered electron, was reconstructed from energy-flow objects
(EFOs)~\cite{epj:c1:81,*thesis:briskin:1998} which combine the
information from calorimetry and tracking, corrected for the energy
loss in the detector material.  
Each EFO was assigned a reconstructed
four-momentum, $q^{i} = (p_X^i,p_Y^i,p_Z^i,E^{i})$.  
Jets were
reconstructed from EFOs using the $k_{T}$ algorithm~\cite{pr:d48:3160}
in the longitudinally invariant mode with
the massive recombination scheme~\cite{np:b406:187}.

The following cuts were applied to select DIS events:
\begin{itemize}
\item the photon virtuality, $Q^{2}$, must be above $10\gev^{2}$,
  where this variable and Bjorken-$x$ were reconstructed using the
  double-angle method~\cite{proc:hera:1991:23};
\item $0.05 < y < 0.7$, where the inelasticity, $y$, was reconstructed
  using the Jacquet-Blondel method~\cite{proc:epfacility:1979:391} for
  the lower cut and the electron method~\cite{proc:hera:1991:23} for
  the higher cut;
\item $40<(E-p_{Z})_{\textrm{tot}}<65\gev$, reconstructed using the
  four-momentum of the final state; this selects fully contained
  neutral-current electron-proton events for which $E-p_{Z}=2\cdot
  E_{e}=55\gev$;
\item $P_{T}/E_{T}<0.7$, where $P_{T}$ and $E_{T}$ are the transverse
  momentum and the scalar transverse energy of the final state. This
  cut was applied to reduce the charged-current and non-$ep$
  backgrounds.
\end{itemize}

In order to estimate the decay length of the $B$ hadron,
a secondary vertex was fitted using all good tracks assigned to the
jet~\cite{thesis:schoenberg:2010}.  Good tracks were defined by a
minimal transverse momentum, $p_{T} > 0.5\gev$, at least four hits in
the MVD and three or more superlayers passed in the CTD.
Vertices with $\chi ^{2}/\dof < 6$ and
a distance from the interaction point within $\pm 1\cm$ in the
$X$--$Y$ plane and $\pm 30\cm$ in the $Z$ direction were taken.

The decay length, $d$, was defined as the distance in $XY$ between the
secondary vertex and the interaction point\footnote{%
  In the $X$--$Y$ plane, the interaction point is defined as the centre of
  the beam ellipse, determined using the average primary vertex
  position for groups of a few thousand events, taking into account the difference in
  angle between the beam direction and the $Z$ direction. The $Z$
  coordinate is taken as the $Z$ position of the primary vertex of
  the event.},
projected onto the jet axis.  
The sign of the decay length was assigned using the axis of the jet to
which the vertex was associated; if the decay length vector 
was in the same hemisphere as the jet axis, a
positive sign was assigned to it, otherwise the sign of the decay
length was negative. Negative decay lengths, which originate from
secondary vertices reconstructed on the wrong side of the interaction
point with respect to the direction of the associated jets, are
unphysical and caused by detector resolution effects. A small
correction~\cite{thesis:shehzadi:2011} to the MC decay-length
distribution was applied in order to reproduce the data with negative
values of decay length; 5\% of the tracks in the central region were
smeared and an additional smearing to tracks in the tails of the
decay-length distribution was applied.

Electron candidates from semileptonic decays of $b$ quarks were
selected from the EFOs having a transverse momentum, $p_{T}^{e}$,
satisfying $0.9 < p_{T}^{e} < 8\gev$ in the pseudorapidity range
$|\eta^{e}| < 1.5$, and consisting of a track matched to a single
calorimetric cluster.  To reduce the hadronic background,
at least $95\%$ of the EFO energy had to be deposited in
the electromagnetic part of the calorimeter.  Candidates in the
angular regions corresponding to the gaps between FCAL and BCAL as well
as between RCAL and BCAL were removed. 
To account for differences in the
$\eta^{e}$ distribution in data and MC, the electron reconstruction efficiency in
MC was corrected by $0.95$ in the FCAL and RCAL regions and by
$1.05$ in the BCAL region.
Electrons from identified photon conversions were rejected~\cite{epj:c18:625}.

The electron candidate was required to be associated
with a jet using the following criteria:
\begin{itemize}
\item the jet was required to have a reconstructed vertex of good quality as defined above;
\item the jet had to have $p_{T}^{\textrm{jet}} > 2.5\gev$ and $|\etajet| < 2.0$;
\item the distance $\Delta R = \sqrt{(\etajet-\eta^{e})^{2} + (\phijet
    - \phi^{e})^{2}} < 1.0$;
\item if there is more than one candidate jet, the jet closest in
  $\Delta R$ to the electron candidate was chosen.
\end{itemize}

The combination of the momentum cut and the jet association reduces
substantially the background from scattered electrons not identified as such.

\begin{figure}[p]
  \vfill\centering
  \includegraphics[width=0.6\textwidth]{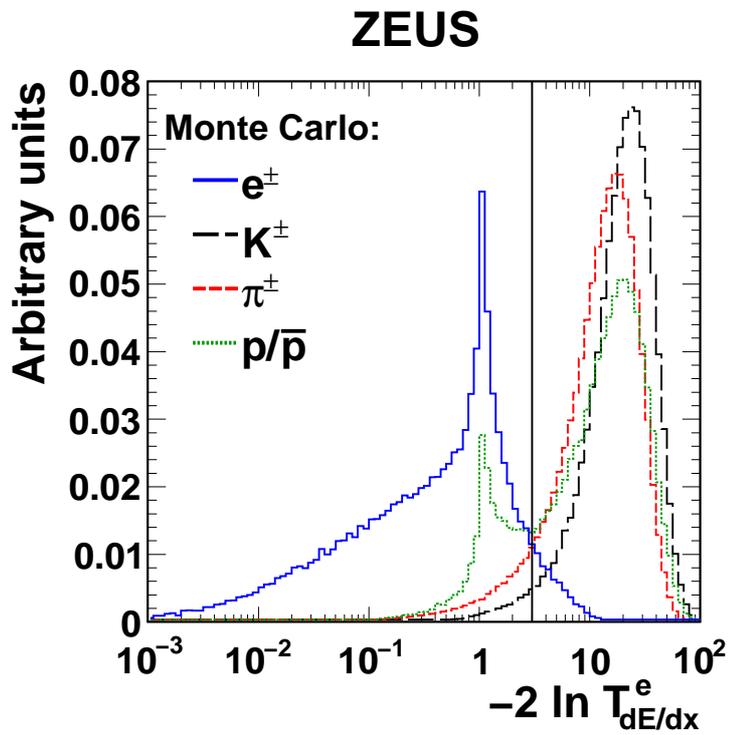}
  \caption{Distribution of the likelihood-ratio test function for the
   electron hypothesis, $T_{\dEdx}^{e}$, for $e^{\pm}, \pi^{\pm},
   K^{\pm}, p$ and $\bar{p}$. All histograms were normalised to
   unity. The selection cut at $- 2 \ln T_{\dEdx}^{e} < 3$ is
   indicated by the vertical line. All other selection cuts were
   applied.}
  \label{fig:dedx}
  \vfill
\end{figure}

The main variable for the electron identification was  
\dEdx{}~\cite{Chekanov:2008aaa}. To reduce the major background of
fake electrons in the candidate selection, a preselection cut was
applied on a likelihood-ratio test function
$T_{e}^{\dEdx}$~\cite{thesis:juengst:2010}.  This function was
calculated using \dEdx{} as discriminating input variable and testing
the electron hypothesis. The distribution of this test function, as
obtained from MC, for the particle types $e^{\pm}, \pi^{\pm},
p/\bar{p}$ and $K^{\pm}$ is shown in Fig.~\ref{fig:dedx}.  
The
vertical line at $-2 \ln T_{e}^{\dEdx}=3$ indicates the cut, which
rejects a large fraction of the background particles.

\section{Identification of electrons from semileptonic decays}
\label{sec:elid}

The electron candidates in the MC samples were classified
into three different categories.  The first category $(b\rightarrow
e)$ contains electrons from beauty decays, including direct
semileptonic decays, cascade decays $b \rightarrow c (\bar{c})
\rightarrow e$, $b \rightarrow \tau \rightarrow e$ and $b \rightarrow
\Jpsi \rightarrow e^{+}e^{-}$.  The second category (other $e$)
contains all true electrons, which are not included in the beauty
signal. These are mainly electrons originating from photon conversions,
Dalitz decays, electrons from direct charm decays, or remaining DIS
electrons.  The third category (non-$e$) includes all candidates
which are fake electrons.  After the selection, the dominant
contribution to the latter comes from pions, while the number of
kaons or protons mimicking electrons is rather small.

For the electron identification, the following three
variables~\cite{Chekanov:2008aaa} were used as discriminants:
\begin{itemize}
\item \dEdx, as measured in the CTD;
\item $E^{\mathrm{CAL}}/p^{\mathrm{track}}$, the energy of the EFO as
  measured in the calorimeter, divided by the track momentum;
\item $d_{\text{cell}}$, the depth of the central energy deposit
within the CAL.
\end{itemize}

The following discriminating variables were used to distinguish the
origin of electron candidates:
\begin{itemize}
\item \pTrel{}, the transverse-momentum component of the electron
  candidate relative to the direction of the jet axis. The shapes of
  the light-quark \pTrel{} distributions in the MC were
  corrected~\cite{thesis:shehzadi:2011} using a background-enriched data
  sample. This variable is sensitive to $b$ decays since electrons
  from $b$ decays tend to have large \pTrel{} due to the large $b$
  mass;
\item \Dphi{}, the difference of azimuthal angles of the electron
  candidate and the missing transverse momentum vector,
  defined as
  \begin{equation*}
    \label{eq:delphi}
    \Dphi = |\phi(\vec{p}_{e})-\phi(\pTmiss)|\,,
  \end{equation*}
  where $\pTmiss$ is the negative vector sum of the EFO momentum transverse to the
  beam axis,
  \begin{equation*}
    \pTmiss = - \textstyle{\left(\sum_{i} p_{x}^{i}, \sum_{i} p_{y}^{i}\right)},
  \end{equation*}
  and the sum runs over all EFOs.  The variable \Dphi{} is sensitive to
  semileptonic decays of $b$ and $c$ hadrons due to the presence of the
  neutrino;
\item $d/\delta d$, the signed decay-length significance, where
  $\delta d$ is the uncertainty on
  $d$~\cite{thesis:yagues:2008,thesis:schoenberg:2010}. This variable
  is sensitive to the decay of $c$ and $b$ hadrons due to their long
  lifetimes.
\end{itemize}
In contrast to the results of a previous ZEUS study~\cite{Chekanov:2008aaa}, the separation power of \Dphi{}
and \pTrel{} is worse due to the lower jet momenta used here.
Therefore it was not possible to
separate the charm signal from the other particles in the electron background. 

Following the procedure of the previous
study~\cite{Chekanov:2008aaa}, the six variables were combined into
one discriminating test-function variable, which is a ratio of
likelihoods. For a given hypothesis of particle, $i$, and source $j$,
the likelihood, $\mathcal{L}_{ij}$, is given by
\begin{equation*}
  \mathcal{L}_{ij} = \prod\limits_{l}\,\mathcal{P}_{ij}(d_{l})\,,
  \label{eq:partid_likelihood}
\end{equation*}
where $\mathcal{P}_{ij}(d_{l})$ is the probability to observe particle
$i$ from source $j$ with value $d_{l}$ of a discriminant variable.
The particle hypotheses $i \in \{e,\pi,K,p\}$ and the sources, $j \in
\{b \shortrightarrow e, \text{other}\,e, \text{non-}e\}$, were considered.
For the likelihood ratio
test, the test function $T_{ij}$ was defined as
\begin{equation*}
  T_{ij} = \frac{\alpha_{i} \alpha'_{j} \mathcal{L}_{ij}}{%
    \sum \limits_{k,l} \alpha_{k} \alpha'_{l} \mathcal{L}_{kl}}.
  \label{eq:partid_testfunc}
\end{equation*}
The $\alpha_{i}$, $\alpha'_{j}$ denote the prior probabilities taken
from MC. In the sum, $k,l$ run over all
particle types and sources defined above. 
In the following, $T$ is always taken to be the likelihood ratio for
an electron originating from a semileptonic $b$-quark decay, $T \equiv
T_{e,\btoe}$, unless otherwise stated.  

\section{Signal extraction}
\label{sec:signal}

The combined MC sample was split into the three contributions
as defined in the previous section. The beauty test function, $T$, was
calculated separately for these three samples and for the data.  The
relative contributions of the three sources in the data,
$f_{\btoe}^{\text{DATA}}$, $f_{\othe}^{\text{DATA}}$,
$f_{\ebkg}^{\text{DATA}}$, were obtained from a three-component
maximum-likelihood fit~\cite{barlow:93} to the $T$ distributions.  The
fit range of the test function was restricted to $-2 \ln T < 10$ to
remove the region dominated by background and where the test function
falls rapidly.  The $\chi^{2}$ for the fit is $\chi^{2}/\mathrm{ndf} =
18/28$.

\begin{figure}[p]
  \vfill\centering
  \includegraphics[width=0.78\textwidth]{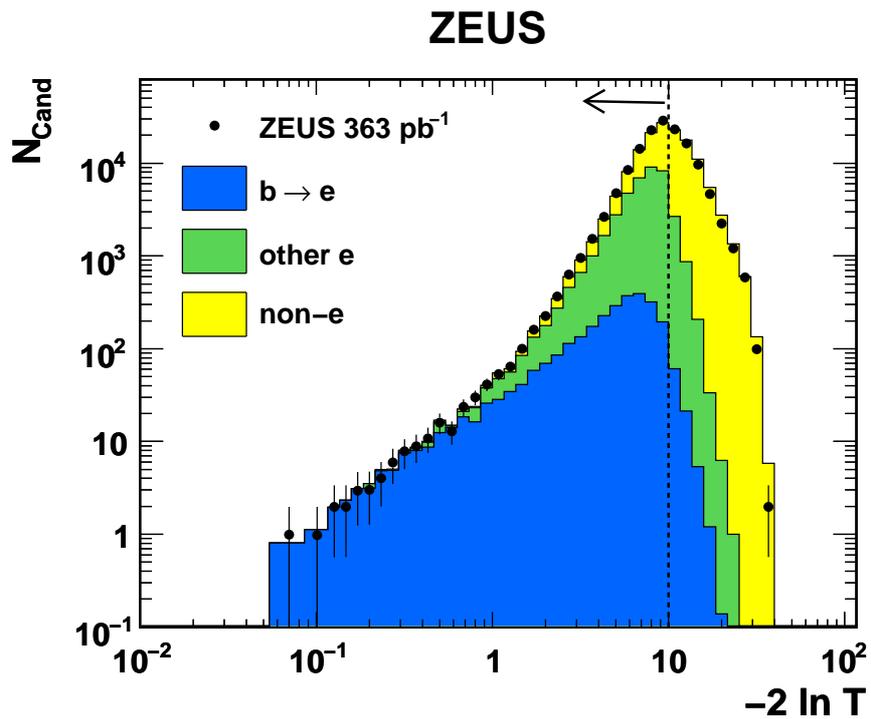}
  \caption{The distribution of $-2 \ln T$, where T is the test
    function, using the beauty hypothesis for electron candidates,
    compared to the Monte Carlo expectation after the fit described in
    the text.  The arrow indicates the region included in the fit ($-2
    \ln T < 10)$.  The shaded areas show the fitted contributions for
    electrons from $b$-quark decays, electrons from other sources and the
    non-electron background.}
  \label{fig:likel_fit}
  \vfill
\end{figure}

\begin{figure}[p]
  \vfill\centering
  \includegraphics[width=0.74\textwidth]{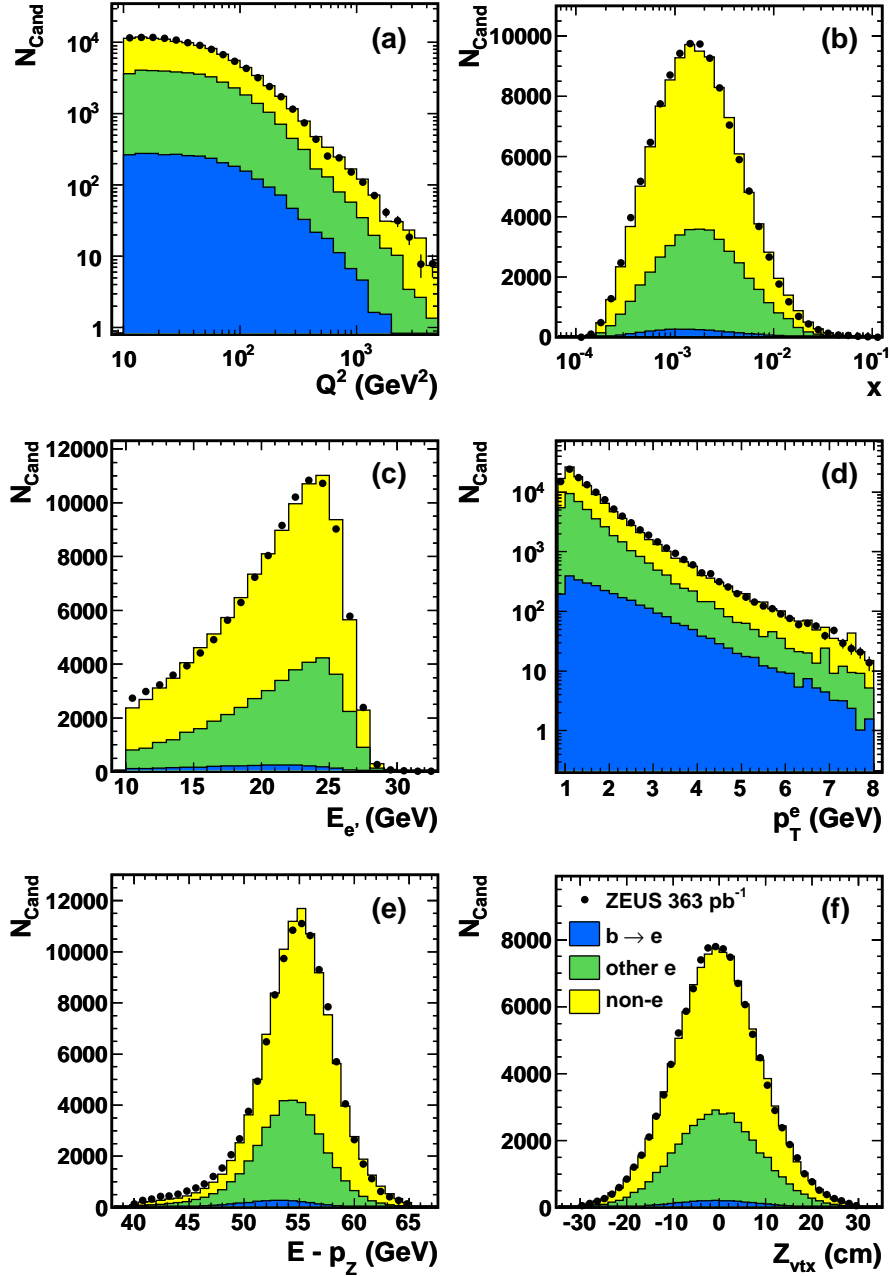}
  \caption{Distributions of the variables related to the event
    selection, after applying all selection cuts: for the kinematic
    variables (a) $Q^{2}$ and (b) $x$, for (c) the energy of the
    scattered electron, $E_{e'}$ and (d) the transverse momentum of
    the electron candidate, $p_{T}^{e}$. The variables $E-p_{Z}$ and
    $Z_{\text{vtx}}$, which were used for the event selection are
    shown in (e) and (f), respectively. The shaded areas show the MC
    expectations for the contributions for electrons from $b$-quark
    decays, electrons from other sources and the non-electron
    background as denoted in the figure, after applying the scale
    factors from the fit. The summed distribution is compared with the
    data distribution shown by the black points.}
  \label{fig:ctrl}
  \vfill
\end{figure}

\begin{figure}[p]
  \vfill\centering
  \includegraphics[width=0.75\textwidth]{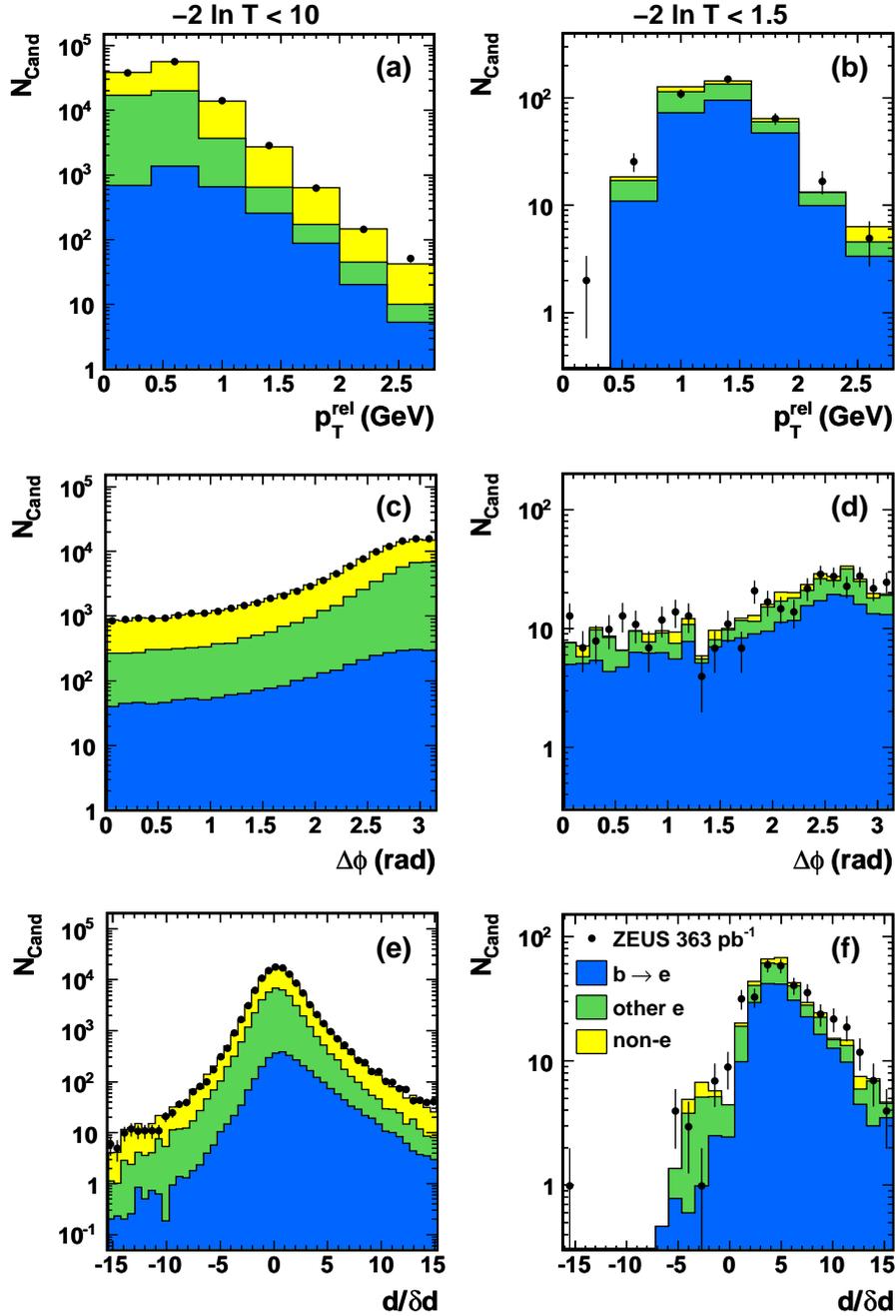}\\
  \caption{Distributions of (a) \pTrel{}, (c) \Dphi{} and (e) \DLsig{}
    for all candidates that enter the fit satisfying $-2 \ln T
    <10$. The same plots are shown for the beauty-enriched region ($-2
    \ln T <1.5$) in (b), (d) and (f). For details see the caption of
    Fig.~\protect\ref{fig:ctrl}.}
    \label{fig:ctrl_likel}
    \vfill
\end{figure}

The result of the fit is shown in Fig.~\ref{fig:likel_fit} and
corresponds to a scaling of the cross section predicted by the beauty
MC by a factor of $1.32 \pm 0.11$. For the other two samples
the scaling factors were determined to be $\sim$1.1 for the electron
background and $\sim$1.3 for the non-$e$ background.  These factors
were applied to the contributions shown in Figs.~\ref{fig:ctrl}
and~\ref{fig:ctrl_likel}.

Figure~\ref{fig:ctrl} shows a comparison of the MC simulation to the
data for the main variables used for the event selection. 
The Monte Carlo describes the data well.
Figure~\ref{fig:ctrl_likel} shows the distributions for the variables
in the likelihood-ratio test function, which are sensitive to the
different origin of the electron candidates. In
Figs.~\ref{fig:ctrl_likel} (a), (c) and (e), the three variables are
shown for the selection used in the fit.
Figures~\ref{fig:ctrl_likel} (b), (d) and (f),
show the same distributions for a signal-enriched region, which
is defined by a harder cut on the test function at $-2 \ln T <
1.5$. All distributions are reasonably well described.

\section{Cross-section determination}
\label{sec:xsect}

The differential beauty cross section for a variable, $v$, was determined
separately for each bin, $k$, from the relative fractions in the data
obtained from the fit and the acceptance correction,
$\mathcal{A}_{\btoe}^{v_{k}}$, calculated using MC events,
\begin{equation}
  \label{eq:diffacceptance}
  \frac{\dif\sigma_{\btoe}}{\dif v_{k}} = 
  \frac{N^{\mathrm{DATA}} \cdot f_{\btoe}^{\mathrm{DATA}}(v_{k})}{%
    \mathcal{A}_{\btoe}^{v_{k}}
    \cdot \mathcal{L} \cdot \Delta v_{k}}\cdot {C}_{r},
\end{equation}
where $N^{\mathrm{DATA}}$ is the number of electron candidates found
in the data bin, $\mathcal{L}$ is the integrated luminosity, $\Delta
v_{k}$ is the bin width and $C_{r}$ is the QED radiative-correction
factor. The acceptance is defined as
\begin{equation*}
  \mathcal{A}_{\btoe} = \frac{N_{\btoe}^{\text{rec}}}{N_{\btoe}^{\text{true}}} ,
\end{equation*}
where $N_{\btoe}^{\textrm{rec}}$ is the number of electrons from
semileptonic decays reconstructed in the MC sample satisfying
the selection criteria detailed in Section~\ref{sec:data}, and
$N_{\btoe}^{\mathrm{true}}$ is the number of electrons from semileptonic
decays produced in the signal process that satisfy the kinematic
requirements of the cross-section definition using the MC
information at the generator level.
The kinematic variables $Q^{2}$ and $x$ at the true level were calculated
using the four-momentum of the exchanged photon after possible
initial-state radiation (ISR).

The cross sections were corrected to the QED Born level, calculated
using a running coupling constant, $\alpha_{em}$, such that they can
be compared directly to the NLO QCD predictions by HVQDIS. The
radiative corrections were obtained using the \RAPGAP{} Monte Carlo as
$C_{r}=\sigma_{\text{Born}}/\sigma_{\text{rad}}$, where $\sigma_{\text{rad}}$
is the cross section with full QED corrections (as used in the
standard MC samples) and $\sigma_{\text{Born}}$ was obtained with the QED
corrections turned off.  The corrections are typically $C_{r}\approx
1.05$ rising to $C_{r}\approx 1.10$ for the high $Q^{2}$ region.

\section{Systematic uncertainties}
\label{sec:syst}
The systematic uncertainties were calculated by varying the analysis
procedure and then repeating the fit to the likelihood
distributions~\cite{thesis:shehzadi:2011}.  
The variations were made in a range such that the MC continued
to provide a reasonable description of the data for the relevant
distributions.
The systematic
uncertainties were determined bin by bin, unless stated otherwise.
The main contributions came from the following sources, where the
numbers in parentheses correspond to the uncertainty on the total cross
section:

\begin{enumerate}
\item DIS selection -- the preselection cuts on the scattered electron
  were varied in both data and MC. The only cuts that had a
  significant effect were the cut on the energy, which was varied
  between $9 < E_{e'} < 11\gev$, the cut on the inelasticity, which was varied 
  between $0.04 < y_{\text{JB}} < 0.06$, and the energy window for $E-p_{Z}$,
  which was varied by $\pm 4\gev$ ($^{+1.7}_{-1.5}{\%}$);
\item trigger efficiency -- the uncertainty on the trigger efficiency
  was evaluated by comparing events taken with independent triggers
  (${+1.2}{\%}$);
\item \dEdx{} simulation -- both the mean and the width of the \dEdx{}
  distribution were varied in the MC separately and
  simultaneously by the uncertainty estimated from the
  data~\cite{thesis:bartsch:2007}. These two variations were then combined,
  giving a conservative estimate of the uncertainty on the \dEdx{}
  test function ($^{+0.4}_{-0.4}{\%}$);
\item tracking efficiency -- the track-finding inefficiency in the data
  with respect to the MC was estimated to be at most 2\%.  The overall
  uncertainty due to this tracking inefficiency was determined by
  randomly rejecting 2\% of all tracks in the MC and repeating the
  secondary-vertex finding (${-3.4}{\%}$);
\item decay-length smearing -- the fraction of events in the MC where
  the decay-length smearing was applied was varied by $\pm 2{\%}$ and
  the additional terms for the smearing of the tails were switched off
  ($^{+2.6}_{-2.0}{\%}$);
\item \pTrel{} shape correction -- the correction applied to the MC was
  switched off and increased by an additional $50\%$
  ($^{-1.5}_{-2.4}{\%}$);
\item electron background -- the relative contributions of the different
  electron sources in the MC were changed by varying separately the
  contributions from photon conversions, Dalitz decays, semileptonic
  decays from charm and DIS electrons by $\pm 25\%$
  ($^{+2.5}_{-2.4}{\%}$);
\item charm-spectrum reweighting -- the correction to the $c$-decay
  electron spectrum in the MC using the CLEO data was varied by $\pm
  50{\%}$ ($^{+3.4}_{-2.9}{\%}$);
\item energy scale -- the global energy scale was varied in the MC by
$\mp 2\%$ ($^{+1.2}_{-1.0}{\%}$);
\item jet energy scale -- the calorimetric part of the transverse jet
  energy in MC was varied by $\pm 3\%$
  ($^{+1.7}_{+0.7}{\%}$);
\item MC model dependence -- the $Q^{2}$ reweighting correction was
  varied by a factor of two ($^{+2.0}_{-1.9}{\%}$);
\item electron reconstruction efficiency -- the electron reconstruction efficiency in MC was varied by
 $\pm0.05$ in the FCAL and RCAL regions and by
$\mp0.05$ in the BCAL region ($^{+4.0}_{-3.7}{\%}$).
\end{enumerate}

A series of further checks were made. 
The fit range was varied to check
possible deficits in the background description. Selection cuts
such as the $Z$ vertex position or preselection cuts such as on the \dEdx{}
test function were varied before repeating the analysis. Another
important check was the charge dependence. Separate fits were made for
electron and positron candidates for each lepton-beam charge
separately as well as for the combined sample.  All variations were
found to be small and consistent with the expected fluctuations due to
statistics and were therefore not included in the systematic
error.

The individual contributions to the systematic uncertainties were
added in quadrature, separately for the negative and the positive
variations, to determine the systematic uncertainty of
$^{+7.4}_{-7.7}\%$ for the total cross section.  A $\pm 2.0\%$ overall
normalisation uncertainty associated with the luminosity measurement
was included in the uncertainty on the total cross section.

\section{Results}
\label{sec:results}

\begin{figure}[p]
  \vfill\centering
  \includegraphics[width=0.8\textwidth]{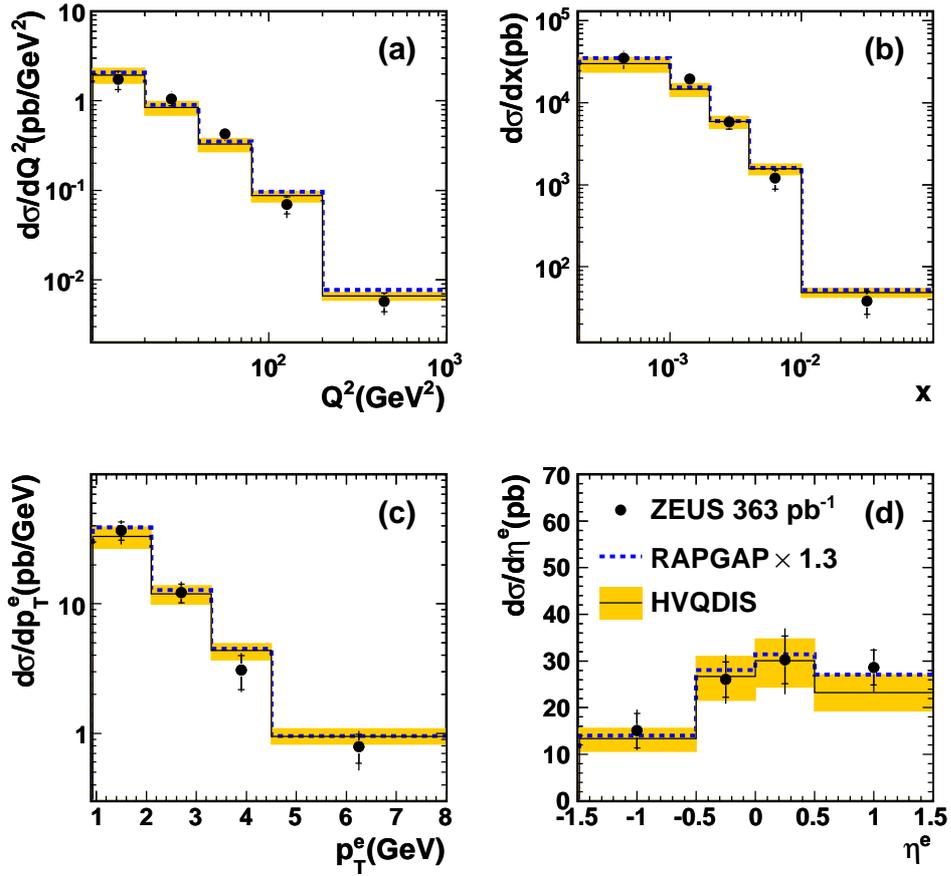}
  \caption{Differential cross sections for electrons from $b$-quark
    decays as a function of the kinematic variables (a) $Q^{2}$ and
    (b) $x$, and the decay electron variables (c) $p_{T}^{e}$ and (d)
    $\eta^{e}$. The cross sections are given for $\Qsq > 10\gev^{2}$,
    $0.05 < y < 0.7$, $0.9 < p_{T}^{e} < 8\gev$ and $|\eta^{e}| < 1.5$. The
    measurements are shown as points.  The inner error bar shows the
    statistical uncertainty and the outer error bar shows the
    statistical and systematic uncertainties added in quadrature.  The
    solid line shows the NLO QCD prediction, with the
    uncertainties indicated by the band; the dashed line shows the
    scaled prediction from \RAPGAP{}.}
  \label{fig:diffxsec}
  \vfill
\end{figure}

\begin{figure}[p]
  \vfill\centering
  \includegraphics[width=0.8\textwidth]{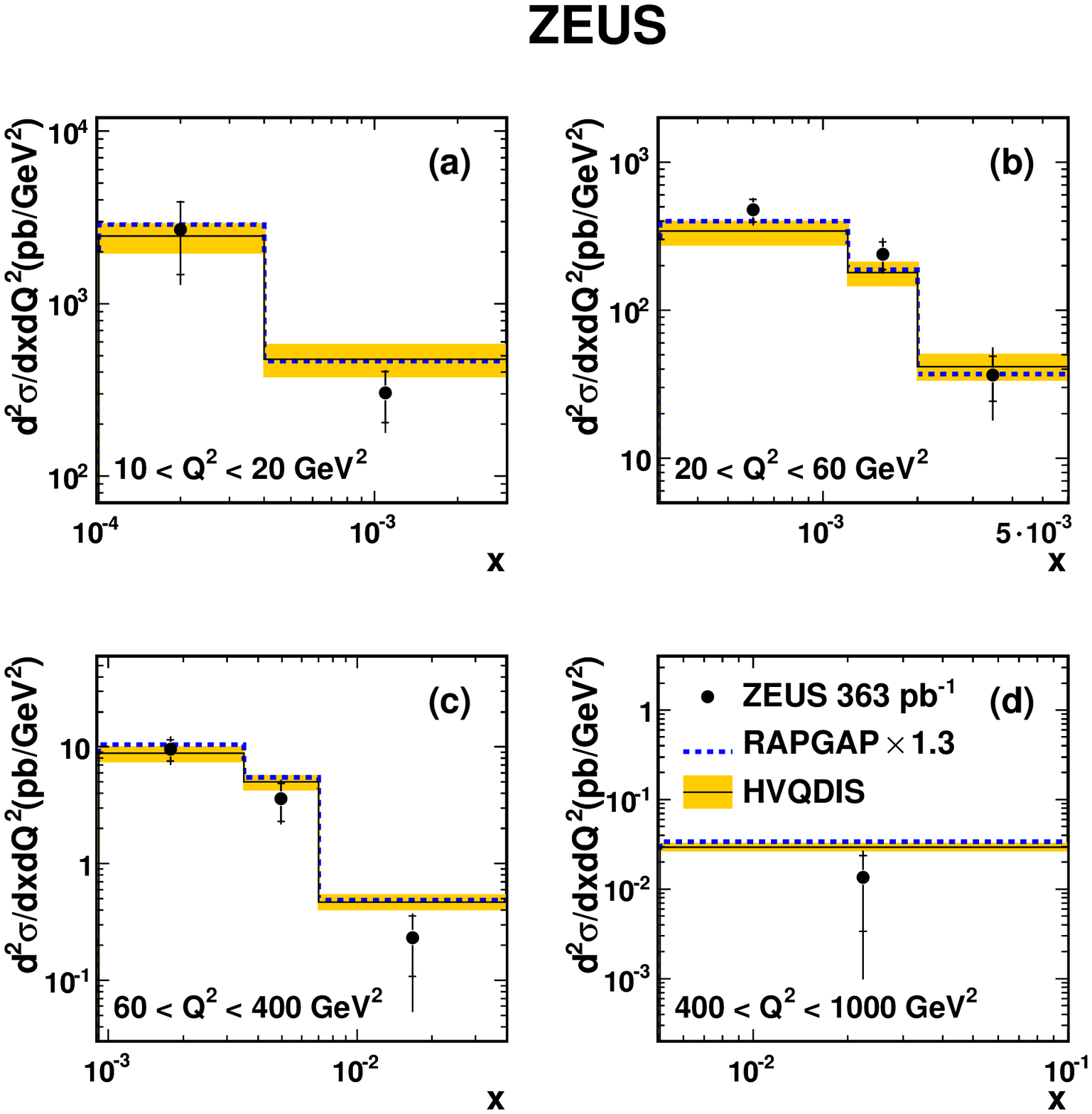}
  \caption{Double-differential cross sections for electrons from $b$-quark 
    decays as a function of 
    $x$ for different regions of $Q^{2}$. 
    Other details as in the caption of
    Fig.~\protect\ref{fig:diffxsec}.  }
  \label{fig:ddiffxsec}
  \vfill
\end{figure}

The visible cross section for electrons from direct and indirect
$b$-quark decays with $0.9 < \pTe < 8\gev$ in the range $|\etae| <
1.5$ was measured in DIS events with $\Qsq > 10 \gev^{2}$ and $0.05 <
y < 0.7$ and found to be
\begin{align*}
\sigma_{\btoe} & = 
\left( 71.8 \pm 5.5 (\stat) ^{+5.3}_{-5.5} (\syst) \right)\pb.
\end{align*}
This cross section includes all electrons and positrons from both $b$
and $\bar{b}$ and no jet requirement was applied at the true level.
This result can be compared to the HVQDIS NLO QCD prediction of
\begin{align*}
\sigma_{\btoe}^{\text{NLO}} & = 
\left( 67 ^{+10}_{-11} \right)\pb,
\end{align*}
where the uncertainty is calculated as described in Section~\ref{sec:theory}.
This value agrees well with the measured cross
section, which is a factor 1.3 higher
than the \RAPGAP{} leading-order prediction\footnote{Note that the
\RAPGAP{} predictions do not include the $Q^{2}$ reweighting
correction discussed in Section~\ref{sec:montecarlo}.}
 of $54.4\pb$. This
factor is used to scale the \RAPGAP{} predictions in
Figs.~\ref{fig:diffxsec} and~\ref{fig:ddiffxsec}.

Differential cross sections as a function of
\pTe{} and $\eta_{e}$, $Q^{2}$ and $x$ are shown in
Fig.~\ref{fig:diffxsec}.  Figure~\ref{fig:ddiffxsec} shows the
differential cross sections as a function of $x$, split into four
different $Q^{2}$ ranges. The figures also
show the NLO QCD and the scaled \RAPGAP{} predictions.
The cross-section values are
given in Tables~\ref{tab:diff1}--\ref{tab:double}.  
Both the predictions from the NLO QCD calculations as well as the scaled
\RAPGAP{} cross sections describe the data well.

\section{\boldmath Extraction of $F_{2}^{b\bar{b}}$}

The structure function $F_{2}^{b\bar{b}}$
can be defined in terms of the inclusive double-differential cross
section (defined in analogy to Eq.~\ref{eq:diffacceptance}) as a
function of $x$ and $Q^{2}$,
\begin{equation*}
\frac{\dif^{2}\sigma_{b\bar{b}}}{\dif x\,\dif Q^{2}}=
\frac{Y_{+}(2\pi \alpha^{2}_{\text{em}})}{xQ^{4}}\left[F_{2}^{b\bar{b}}(x,Q^{2})-\frac{y^{2}}{Y_{+}}F_{L}^{b\bar{b}}(x
,Q^{2})\right]\,,
\end{equation*}
where $Y_{+}=1+(1-y)^{2}$ and $F_{L}^{b\bar{b}}$ is the beauty
contribution to the structure function $F_{L}$.

The electron cross section, $\sigma_{b \shortrightarrow e}$, measured
in bins of $x$ and $Q^{2}$, was used to extract $F_{2}^{b\bar{b}}$ at
a reference point in the $x$--$Q^{2}$ plane using
\begin{equation*}
  F_{2}^{b\bar{b}}(x,Q^{2}) =
  \frac{\dif^{2}\sigma_{\btoe}}{\dif x\,\dif Q^{2}} \cdot
  \frac{F_{2}^{b\bar{b},\text{NLO}}(x,Q^{2})}%
  {\dif^{2}\sigma_{\btoe}^{\text{NLO}}/\dif x\,\dif Q^{2}}\,,
\end{equation*}
where $F_{2}^{b\bar{b},\text{NLO}}$ and
$\dif^{2}\sigma_{\btoe}^{\text{NLO}}/\dif x\,\dif Q^{2}$
were calculated in the FFNS using the HVQDIS program. The uncertainty
on the extrapolation from the measured range to the full kinematic
phase space was estimated by varying the settings of the calculation
(see Section~\ref{sec:theory}) for $F_{2}^{b\bar{b},\text{NLO}} /
(\dif^{2}\sigma_{\btoe}^{\text{NLO}}/\dif x\,\dif Q^{2})$ and adding the resulting
uncertainties in quadrature.  For each bin, a reference point in $x$
and $Q^{2}$ was defined (see Table~\ref{tab:f2b}) to calculate the
structure function. The small correction for $F_{L}^{b\bar{b}}$ is taken into
account in the HVQDIS prediction.

\begin{figure}[p]
  \vfill\centering
  \includegraphics[width=0.9\textwidth]{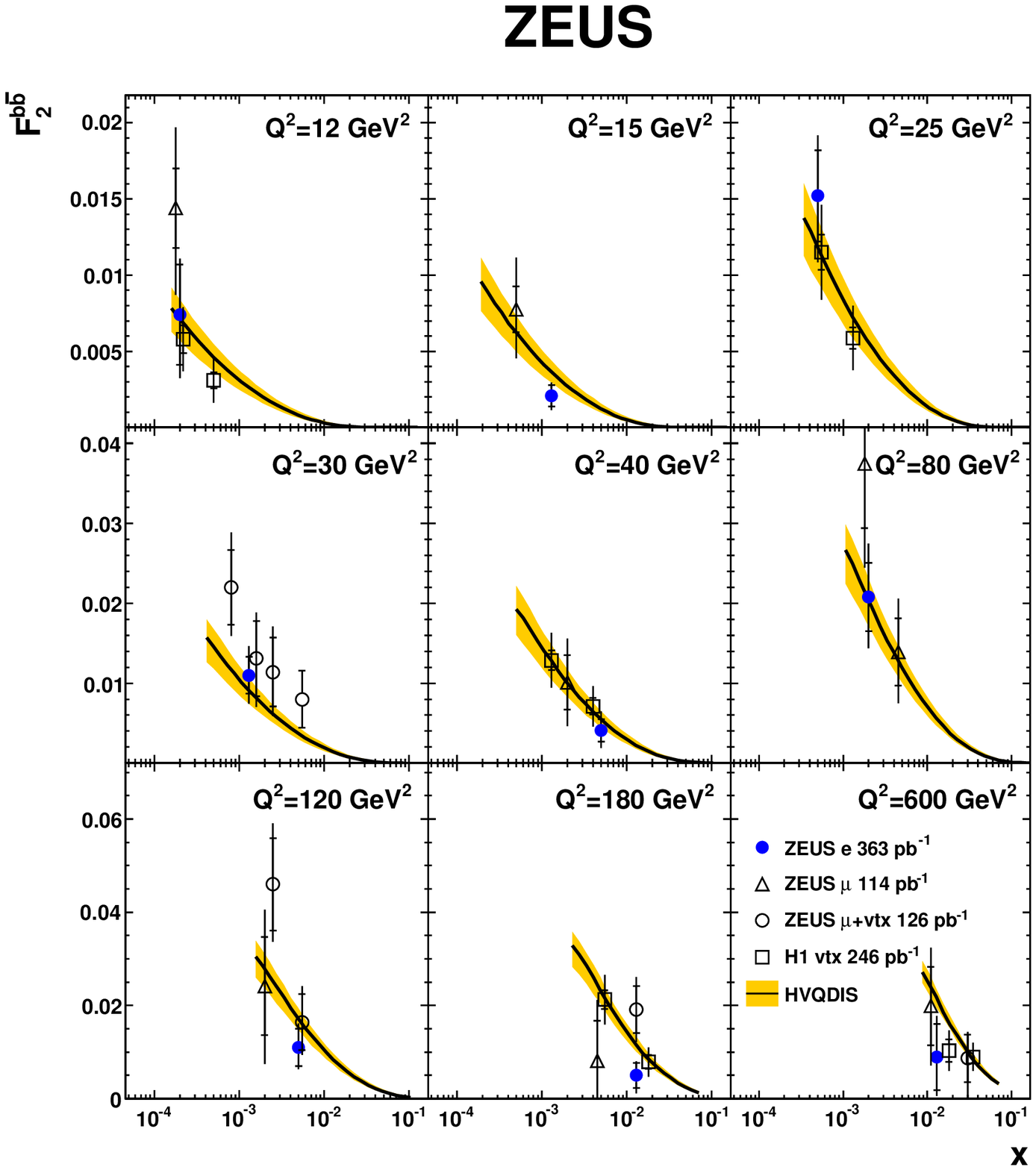}
  \caption{The structure function $F_{2}^{b\bar{b}}$ (filled symbols)
    as a function of $x$ for nine different values of $Q^{2}$ compared
    to previous results (open symbols). The inner error bars are the
    statistical uncertainty while the outer error bars represent the
    statistical, systematic and extrapolation uncertainties added in
    quadrature. The band represents the uncertainty on the NLO QCD
    prediction. Previous data have been corrected to the reference
    $Q^{2}$ range of this analysis.}
  \label{fig:f2b-a}
  \vfill
\end{figure}

The structure function $F_{2}^{b\bar{b}}$ is shown in
Fig.~\ref{fig:f2b-a} as a function of $x$ for nine different values of
$Q^{2}$. The values and the corresponding uncertainties are given in
Table~\ref{tab:f2b}. To compare the result with previous
measurements~\cite{Abramowicz:2010zq,Chekanov:2009kj,:2009ut}, the
earlier results were extrapolated to the $Q^{2}$ values chosen in this
analysis.
For $Q^{2} > 10\gev^{2}$, this measurement represents the most precise
determination of $F_{2}^{b\bar{b}}$ by the ZEUS Collaboration.  It is
in good agreement with previous ZEUS analyses and the H1
measurement. The NLO QCD prediction describes the data well.  The same
measurements are also shown as a function of $Q^{2}$ for fixed $x$
in Fig.~\ref{fig:f2b-b}, compared to several NLO and NNLO QCD
predictions based on the fixed- or variable-flavour-number
schemes~\cite{PhysRevD.78.013004,*springerlink:10.1140/epjc/s10052-009-1072-5,*Thorne:2008xf,*glueck:2008,*Alekhin2009166,*Alekhin:2009ni,*Alekhin:2010iu}. For
the HVQDIS prediction shown in this figure, the scale parametrisation
$\mu=\frac{1}{2}\sqrt{Q^{2}+p_{T}^{2}+m_{b}^{2}}$~\cite{Abramowicz:2010zq}, was used. All the theoretical predictions shown
provide a good description of the data.

\section{Conclusions}
\label{sec:conclusions}

Beauty production has been measured in DIS using semileptonic decays
into electrons. A likelihood-ratio test function, adapted from a previous
measurement, was used to identify the signal. The analysis benefited from the
improved tracking in the HERA II data-set through the use of the
measured decay length of weakly decaying $B$ hadrons.

The total cross section and differential cross
sections as a function of $x$, $Q^{2}$, $p_{T}^{e}$ and $\eta^{e}$
were determined.  NLO QCD predictions calculated using the HVQDIS
program describe the data well.  The \RAPGAP{} Monte Carlo provides a
good description of the shape of the differential distributions.

The structure function  $F^{b\bar{b}}_{2}$ was extracted from
the double-differential cross section as a function of $x$ and $Q^{2}$.
The measurement is in agreement with the results obtained from
previous analyses using different techniques.
For $Q^{2} > 10\gev^{2}$, this measurement represents the most precise
determination of $F_{2}^{b\bar{b}}$ by the ZEUS Collaboration.
The results were also compared to several NLO and NNLO QCD
calculations, which provide a good description of the data.

\section*{Acknowledgements}
\label{sec:acknowledge}

\begin{figure}[p]
  \vfill\centering
  \includegraphics[width=0.8\textwidth]{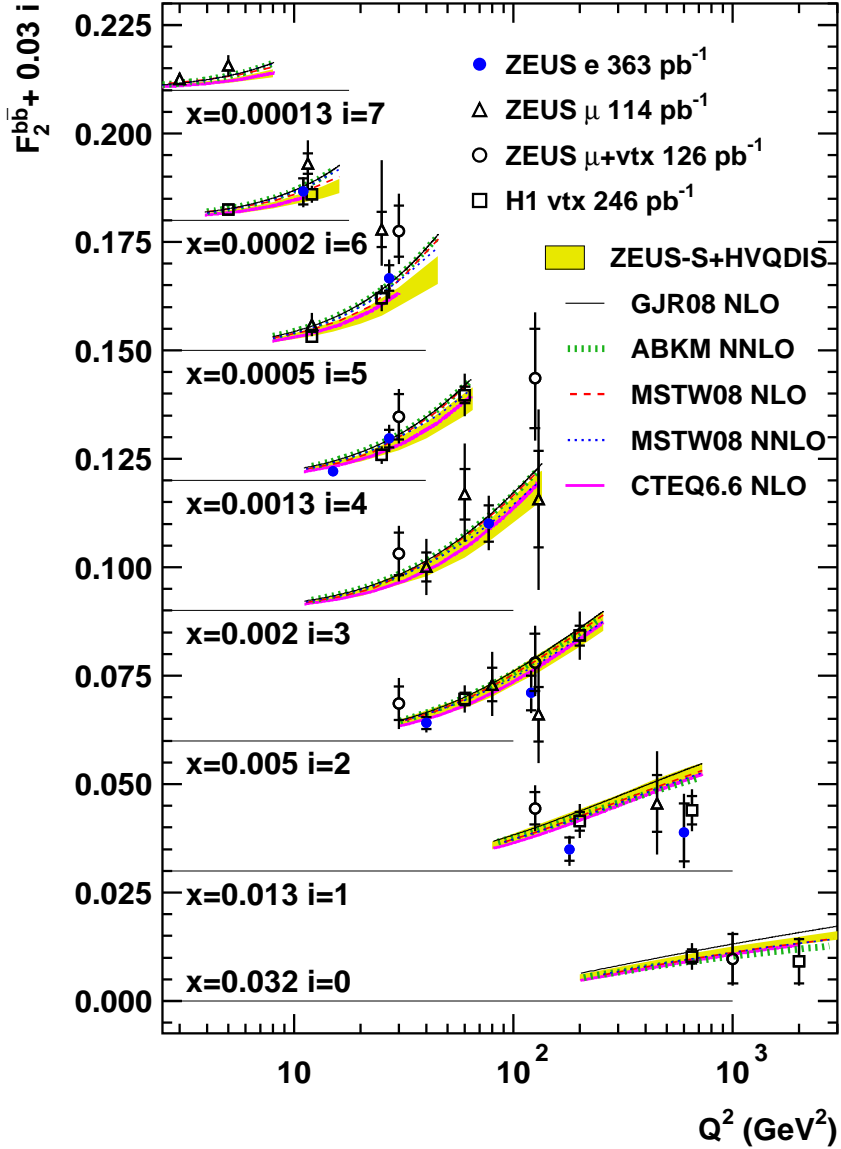}
  \caption{The structure function $F_{2}^{b\bar{b}}$ (filled symbols)
    as a function of $Q^{2}$ for fixed values of $x$ compared to
    previous results (open symbols). The inner error bars are the statistical uncertainty
    while the outer error bars represent the statistical, systematic
    and extrapolation uncertainties added in quadrature.
    The data have been corrected to the same reference $x$ as the
    previous analysis~\protect\cite{Abramowicz:2010zq}.  The measurements are compared to several
    NLO and NNLO QCD predictions (see text for details).}
  \label{fig:f2b-b}
  \vfill
\end{figure}

It is a pleasure to thank the ABKM, CTEQ, GJR and MRST groups that
provided the predictions for $F^{b\bar{b}}_{2}$ shown in
Fig.~\ref{fig:f2b-b}.
We appreciate the contributions to the construction and maintenance of
the ZEUS detector of many people who are not listed as authors.  The
HERA machine group and the DESY computing staff are especially
acknowledged for their success in providing excellent operation of the
collider and the data-analysis environment. We thank the DESY
directorate for their strong support and encouragement.

\vfill\eject


\clearpage
{\raggedright
\providecommand{\etal}{et al.\xspace}
\providecommand{\coll}{Collab.\xspace}
\catcode`\@=11
\def\@bibitem#1{%
\ifmc@bstsupport
  \mc@iftail{#1}%
    {;\newline\ignorespaces}%
    {\ifmc@first\else.\fi\orig@bibitem{#1}}
  \mc@firstfalse
\else
  \mc@iftail{#1}%
    {\ignorespaces}%
    {\orig@bibitem{#1}}%
\fi}%
\catcode`\@=12

}
\clearpage
%
\newcommand{\phd}{\phantom{.}}
\newcommand{\pho}{\phantom{0}}

\begin{table}[p]
  \centering
  \begin{tabular}{|r@{~:~}l  | r@{$\pm$}>{$}l<{$} r@{$^{+}_{-}$}>{$}l<{$}|}
    \hline
    \multicolumn{2}{|c|}{\centering $Q^{2}$} &
    \multicolumn{2}{c}{\centering $\dif\sigma_{\btoe}/\dif Q^{2}$} &
    \multicolumn{2}{c|}{\centering $\dif\signlob_{\btoe}/\dif Q^{2}$}\\[-0.2em]
    
    \multicolumn{2}{|c|}{(\gev$^{2}$)}&
    \multicolumn{2}{c}{(pb/\gev$^{2}$)}&
    \multicolumn{2}{c|}{(pb/\gev$^{2}$)}\\ 
    \hline
    10&20 &  1.73    & 0.40^{+0.20}_{-0.29}   &\quad 1.93  & ^{0.37}_{0.37} \\
    20&40 &  1.05    & 0.18^{+0.12}_{-0.07}   & 0.84  & ^{0.13}_{0.15} \\
    40&80 &  0.428    & 0.063^{+0.036}_{-0.037}   & 0.327  & ^{0.050}_{0.057} \\
    80&200 &  0.070    & 0.015^{+0.006}_{-0.014}   & 0.087  & ^{0.011}_{0.013} \\
    200&1000 &  0.0057    & 0.0014^{+0.0003}_{-0.0010}   & 0.0066  & ^{0.0006}_{0.0007} \\
    \hline
    \multicolumn{6}{c}{}\\
    \hline
    \multicolumn{2}{|c|}{\centering $x$} & 
    \multicolumn{2}{c}{\centering $\dif\sigma_{\btoe}/\dif x$} &
    \multicolumn{2}{c|}{\centering $\dif\signlob_{\btoe}/\dif x$} \\[-0.2em]
    
    \multicolumn{2}{|c|}{~} & 
    \multicolumn{2}{c}{(pb)}&
    \multicolumn{2}{c|}{(pb)}\\ 
    
    \hline
    0.0002&0.0010 &  34800    & 5700^{+5400}_{-7300}   &\quad 29700  & ^{5400}_{6100} \\
    0.0010&0.0020 &  19400    & 2700^{+1900}_{-1900}   & 14700  & ^{2400}_{2800} \\
    0.0020&0.0040 &  5800    & 1100^{+600}_{-400}   & 5900  & ^{900}_{1100} \\
    0.0040&0.0100 &  1200    & 310^{+210}_{-220}   & 1560  & ^{220}_{230} \\
    0.0100&0.1000 &  38.4    & 12.1^{+9.7}_{-8.7}   & 48.5  & ^{6.2}_{5.7} \\
    \hline
  \end{tabular}
  
  \caption{Differential cross sections for electrons from $b$-quark
    decays as a function of $Q^{2}$ and $x$. The cross sections are
    given for $\Qsq > 10\gev^{2}$, $0.05 < y < 0.7$, $0.9 < \pTe <
    8\gev$ and $|\etae| < 1.5$.  The first uncertainty is statistical
    and the second is systematic. In addition, the NLO QCD prediction
    and its uncertainty are given.}
  \label{tab:diff1}
\end{table}

\begin{table}[p]
  \centering

  \begin{tabular}{|r@{~:~} l  | r@{$\pm$}>{$}l<{$}
      r@{$^{+}_{-}$}>{$}l<{$} |}
    \hline
    \multicolumn{2}{|c|}{\centering $p_{T}^{e}$} &
    \multicolumn{2}{c}{\centering $\dif\sigma_{\btoe}/\dif p_{T}^{e}$} &
    \multicolumn{2}{c|}{\centering $\dif\signlob_{\btoe}/\dif p_{T}^{e}$}\\[-0.2em]
    
    \multicolumn{2}{|c|}{(\gev)}&
    \multicolumn{2}{c}{(pb/\gev)}&
    \multicolumn{2}{c|}{(pb/\gev)}\\ 
    \hline
    0.9&2.1 &  36.9    & 6.1^{+4.2}_{-5.7}   &\quad 33.1  & ^{6.1}_{6.3} \\
    2.1&3.2 &  12.2    & 2.0^{+1.7}_{-0.8}   & 12.0  & ^{1.8}_{2.0} \\
    3.2&4.5 &  3.08    & 0.90^{+0.60}_{-0.44}   & 4.36  & ^{0.59}_{0.67} \\
    4.5&8.0 &  0.78    & 0.20^{+0.16}_{-0.18}   & 0.95  & ^{0.13}_{0.12} \\
    \hline
    \multicolumn{6}{c}{}\\
    \hline
    \multicolumn{2}{|c|}{\centering $\eta^{e}$} & 
    \multicolumn{2}{c}{\centering $\dif\sigma_{\btoe}/\dif \eta^{e}$} &
    \multicolumn{2}{c|}{\centering $\dif\signlob_{\btoe}/\dif \eta^{e}$} \\[-0.2em]
    
    \multicolumn{2}{|c|}{~} & 
    \multicolumn{2}{c}{(pb)}&
    \multicolumn{2}{c|}{(pb)}\\ 
    
    \hline
    -1.5&-0.5 &  15.1    & 3.7^{+2.7}_{-2.0}   &\quad 13.4  & ^{2.3}_{2.7} \\
    -0.5&0.0 &  26.0    & 3.8^{+3.7}_{-3.6}   & 26.7  & ^{4.3}_{5.1} \\
    0.0&0.5 &  30.3    & 5.1^{+4.4}_{-5.3}   & 30.0  & ^{4.7}_{5.6} \\
    0.5&1.5 &  28.6    & 3.7^{+1.7}_{-3.6}   & 23.2  & ^{3.9}_{3.9} \\
    \hline
  \end{tabular}
  \caption{Differential cross sections for electrons from $b$-quark
    decays as a function of $p_{T}^{e}$ and $\eta^{e}$. The cross
    sections are given for $\Qsq > 10\gev^{2}$, $0.05 < y < 0.7$,
    $0.9 < \pTe < 8\gev$ and $|\etae| < 1.5$.  The first uncertainty
    is statistical and the second is systematic. In addition, the NLO
    QCD prediction and its uncertainty are given.}
  \label{tab:dif2}
\end{table}

\begin{table}[p]
  \centering
  \begin{tabular}{|r@{~:~}l  r@{~:~}l | r@{$\pm$}>{$}l<{$}  
      r@{$^{+}_{-}$}>{$}l<{$}|}
    \hline
    \multicolumn{2}{|c}{$Q^{2}$}&
    \multicolumn{2}{c|}{$x$}&
    \multicolumn{2}{c}{~~$\dif^{2}\sigma_{\btoe}/\dif x\,\dif Q^{2}$~~} &
    \multicolumn{2}{c|}{~~$\dif^{2}\signlob_{\btoe}/\dif x\,\dif Q^{2}$~~} \\ [-0.2em]
    \multicolumn{2}{|c}{(\gev$^{2}$)}&
    \multicolumn{2}{c|}{}&
    \multicolumn{2}{c}{(pb/\gev$^{2}$)}&
    \multicolumn{2}{c|}{(pb/\gev$^{2}$)} \\ 
    \hline
    10&20& 0.0001&0.0004    & 2700 &  1200^{+300}_{-700}   & \quad2500 & ^{400}_{500}\\
    10&20& 0.0004&0.0030    & 300 &  100^{+40}_{-80}   & \quad 480 & ^{100}_{100}\\
    \hline
    20&60& 0.0003&0.0012    & 477 &  84^{+47}_{-60}   & \quad 343 & ^{525}_{650}\\
    20&60& 0.0012&0.0020    & 239 &  51^{+47}_{-36}   & \quad 180 & ^{300}_{325}\\
    20&60& 0.0020&0.0060    & 36 &  12^{+15}_{-14}   & \quad 42 & ^{8}_{8}\\
    \hline
    60&400& 0.0009&0.0035    & 9.6 &  2.0^{+1.9}_{-1.6}   & \quad 8.9 & ^{1.0}_{1.3}\\
    60&400& 0.0035&0.0070    & 3.6 &  1.3^{+1.0}_{-0.6}   & \quad 5.0 & ^{0.6}_{0.7}\\
    60&400& 0.0070&0.0400    & 0.23 &  0.12^{+0.06}_{-0.13}   & \quad 0.47 & ^{0.06}_{0.06}\\
    \hline
    400&1000& 0.0050&0.1000    & 0.013 &  0.010^{+0.009}_{-0.007}   & \quad0.029 & ^{0.002}_{0.003}\\
    \hline
  \end{tabular}
  \caption{Double-differential cross sections for electrons from
    $b$-quark decays as a function of $x$ for four different $Q^{2}$
    ranges. The cross sections are given for $\Qsq > 10\gev^{2}$,
    $0.05 < y < 0.7$, $0.9 < \pTe < 8\gev$ and $|\etae| < 1.5$.  The
    first uncertainty is statistical and the second is systematic. In
    addition, the NLO QCD prediction and its uncertainty are given.}
  \label{tab:double}
\end{table}

\begin{table}[p]
  \centering
  \begin{tabular}{| c c | r @{\,}>{$}l<{$} @{\,}>{$}l<{$} @{\,}>{$}l<{$}|}
    \hline
    $Q^{2}$ (\Gev)& $x$ &
    \multicolumn{4}{c|}{~~$F_{2}^{b\bar{b}}$~~}\\[0.2em]
    \hline
    
     12 & 0.0002& 0.0074 &  \pm 0.0033 & ^{+0.0010}_{-0.0020} & ^{+0.0012}_{-0.0015}\\
     15 & 0.0013& 0.0021 &  \pm 0.0007 & ^{+0.0003}_{-0.0005} & ^{+0.0004}_{-0.0004}\\
     25 & 0.0005& 0.0152 &  \pm 0.0027 & ^{+0.0015}_{-0.0019} & ^{+0.0025}_{-0.0029}\\
     30 & 0.0013& 0.0110 &  \pm 0.0023 & ^{+0.0022}_{-0.0017} & ^{+0.0019}_{-0.0021}\\
     40 & 0.005 & 0.0041 &  \pm 0.0014 & ^{+0.0017}_{-0.0016} & ^{+0.0009}_{-0.0007}\\
     80 & 0.002 & 0.0208 &  \pm 0.0043 & ^{+0.0041}_{-0.0036} & ^{+0.0029}_{-0.0032}\\
    120 & 0.005 & 0.0110 &  \pm 0.0040 & ^{+0.0029}_{-0.0019} & ^{+0.0015}_{-0.0015}\\
    180 & 0.013 & 0.0050 &  \pm 0.0027 & ^{+0.0014}_{-0.0027} & ^{+0.0006}_{-0.0006}\\
    600 & 0.013 & 0.0089 &  \pm 0.0067 & ^{+0.0057}_{-0.0048} & ^{+0.0008}_{-0.0008}\\

    \hline
  \end{tabular}
  \caption{The structure
    function $F_{2}^{b\bar{b}}$ given for nine different values of
    $Q^{2}$ and $x$. The first error is statistical, the
    second systematic and the last is the extrapolation
    uncertainty.}
  \label{tab:f2b}
\end{table}


\clearpage

\clearpage 

\clearpage 

\clearpage 

\clearpage 

\clearpage 

\clearpage 

\clearpage 

\clearpage

%
%

\begin{mcbibliography}{10}

\bibitem{np:b392:162}
E.~Laenen \etal,
\newblock Nucl.\ Phys.{} {\bf B 392},~162~(1993)\relax
\relax
\bibitem{np:b392:229}
E.~Laenen \etal,
\newblock Nucl.\ Phys.{} {\bf B~392},~229~(1993)\relax
\relax
\bibitem{Abramowicz:2010zq}
ZEUS \coll, H.~Abramowicz et. al.,
\newblock Eur. Phys. J.{} {\bf C~69},~347~(2010)\relax
\relax
\bibitem{Chekanov:2009kj}
ZEUS \coll, S.~Chekanov et. al.,
\newblock Eur. Phys. J.{} {\bf C~65},~65~(2010)\relax
\relax
\bibitem{Chekanov:2008tx}
ZEUS \coll, S.~Chekanov et. al.,
\newblock JHEP{} {\bf 04},~133~(2009)\relax
\relax
\bibitem{Chekanov:2008zz}
ZEUS \coll, S.~Chekanov et. al.,
\newblock JHEP{} {\bf 02},~032~(2009)\relax
\relax
\bibitem{Chekanov:2008aaa}
ZEUS \coll, S.~Chekanov et. al.,
\newblock Phys. Rev.{} {\bf D~78},~072001~(2008)\relax
\relax
\bibitem{epj:c50:1434}
ZEUS \coll, S.~Chekanov \etal,
\newblock Eur.\ Phys.\ J.{} {\bf C~50},~1434~(2007)\relax
\relax
\bibitem{pl:b599:173}
ZEUS \coll, S.~Chekanov \etal,
\newblock Phys.\ Lett.{} {\bf B~599},~173~(2004)\relax
\relax
\bibitem{pr:d70:012008}
ZEUS \coll, S.~Chekanov \etal,
\newblock Phys.\ Rev.{} {\bf D~70},~12008~(2004).
\newblock Erratum-ibid {\bf D~74}, 59906 (2006)\relax
\relax
\bibitem{epj:c18:625}
ZEUS \coll, J.~Breitweg \etal,
\newblock Eur.\ Phys.\ J.{} {\bf C~18},~625~(2001)\relax
\relax
\bibitem{:2009ut}
H1 \coll F.D.~Aaron \etal,
\newblock Eur. Phys. J.{} {\bf C~65},~89~(2010)\relax
\relax
\bibitem{epj:c47:597}
H1 \coll, A.~Aktas \etal,
\newblock Eur.\ Phys.\ J.{} {\bf C~47},~597~(2006)\relax
\relax
\bibitem{epj:c45:23}
H1 \coll, A.~Aktas \etal,
\newblock Eur.\ Phys.\ J.{} {\bf C~45},~23~(2006)\relax
\relax
\bibitem{pl:b621:56}
H1 \coll, A.~Aktas \etal,
\newblock Phys.\ Lett.{} {\bf B~621},~56~(2005)\relax
\relax
\bibitem{epj:c40:349}
H1 \coll, A.~Aktas \etal,
\newblock Eur.\ Phys.\ J.{} {\bf C~40},~349~(2005)\relax
\relax
\bibitem{epj:c41:453}
H1 \coll, A.~Aktas \etal,
\newblock Eur.\ Phys.\ J.{} {\bf C~41},~453~(2005)\relax
\relax
\bibitem{pl:b467:156}
H1 \coll, C.~Adloff \etal,
\newblock Phys.\ Lett.{} {\bf B~467},~156~(1999)\relax
\relax
\bibitem{zeus:1993:bluebook}
ZEUS \coll, U.~Holm~(ed.),
\newblock {\em The {ZEUS} Detector}.
\newblock Status Report (unpublished), DESY (1993),
\newblock available on
  \texttt{http://www-zeus.desy.de/bluebook/bluebook.html}\relax
\relax
\bibitem{nim:a279:290}
N.~Harnew \etal,
\newblock Nucl.\ Instr.\ and Meth.{} {\bf A~279},~290~(1989)\relax
\relax
\bibitem{npps:b32:181}
B.~Foster \etal,
\newblock Nucl.\ Phys.\ Proc.\ Suppl.{} {\bf B~32},~181~(1993)\relax
\relax
\bibitem{nim:a338:254}
B.~Foster \etal,
\newblock Nucl.\ Instr.\ and Meth.{} {\bf A~338},~254~(1994)\relax
\relax
\bibitem{nim:a581:656}
A. Polini et al.,
\newblock Nucl.\ Instr.\ and Meth.{} {\bf A~581},~656~(2007)\relax
\relax
\bibitem{pl:b481:213}
ZEUS \coll, J.~Breitweg \etal,
\newblock Phys.\ Lett.{} {\bf B~481},~213~(2000)\relax
\relax
\bibitem{thesis:bartsch:2007}
D.~Bartsch,
\newblock Ph.D. Thesis, Universit\"at Bonn, Bonn, Germany, Report
  \mbox{BONN-IR-2007-05}, 2007,
\newblock available on \texttt{http://brock.physik.uni-bonn.de/zeus\usc
  pub.php}\relax
\relax
\bibitem{nim:a309:77}
M.~Derrick \etal,
\newblock Nucl.\ Instr.\ and Meth.{} {\bf A~309},~77~(1991)\relax
\relax
\bibitem{nim:a309:101}
A.~Andresen \etal,
\newblock Nucl.\ Instr.\ and Meth.{} {\bf A~309},~101~(1991)\relax
\relax
\bibitem{nim:a321:356}
A.~Caldwell \etal,
\newblock Nucl.\ Instr.\ and Meth.{} {\bf A~321},~356~(1992)\relax
\relax
\bibitem{nim:a336:23}
A.~Bernstein \etal,
\newblock Nucl.\ Instr.\ and Meth.{} {\bf A~336},~23~(1993)\relax
\relax
\bibitem{desy-92-066}
J.~Andruszk\'ow \etal,
\newblock Preprint \mbox{DESY-92-066}, DESY, 1992\relax
\relax
\bibitem{zfp:c63:391}
ZEUS \coll, M.~Derrick \etal,
\newblock Z.\ Phys.{} {\bf C~63},~391~(1994)\relax
\relax
\bibitem{acpp:b32:2025}
J.~Andruszk\'ow \etal,
\newblock Acta Phys.\ Pol.{} {\bf B~32},~2025~(2001)\relax
\relax
\bibitem{physics-0512153}
M.~Helbich \etal,
\newblock Nucl.\ Instr.\ and Meth.{} {\bf A~565},~572~(2006)\relax
\relax
\bibitem{cpc:86:147}
H.~Jung,
\newblock Comp.\ Phys.\ Comm.{} {\bf 86},~147~(1995).
\newblock See also \texttt{http://projects.hepforge.org/rapgap/}\relax
\relax
\bibitem{epj:c12:375}
CTEQ \coll, H.L.~Lai \etal,
\newblock Eur.\ Phys.\ J.{} {\bf C~12},~375~(2000)\relax
\relax
\bibitem{cpc:69:155}
A.~Kwiatkowski, H.~Spiesberger and H.-J.~M\"ohring,
\newblock Comp.\ Phys.\ Comm.{} {\bf 69},~155~(1992).
\newblock Also in {\it Proc.\ Workshop Physics at HERA}, eds. W.~Buchm\"{u}ller
  and G.Ingelman, (DESY, Hamburg, 1991)\relax
\relax
\bibitem{proc:hera:1991:1419}
G.A.~Schuler and H.~Spiesberger,
\newblock {\em Proc.\ Workshop on Physics at {HERA}}, W.~Buchm\"uller and
  G.~Ingelman~(eds.), Vol.~3, p.~1419.
\newblock Hamburg, Germany, DESY (1991)\relax
\relax
\bibitem{cpc:71:15}
L.~L\"onnblad,
\newblock Comp.\ Phys.\ Comm.{} {\bf 71},~15~(1992)\relax
\relax
\bibitem{physlett:b667:1}
Particle Data Group, C.~Amsler \etal,
\newblock Phys.~Lett.{} {\bf B~667},~1~(2008)\relax
\relax
\bibitem{cpc:82:74}
T.~Sj\"ostrand,
\newblock Comp.\ Phys.\ Comm.{} {\bf 82},~74~(1994)\relax
\relax
\bibitem{cpc:135:238}
T.~Sj\"{o}strand \etal,
\newblock Comp.\ Phys.\ Comm.{} {\bf 135},~238~(2001)\relax
\relax
\bibitem{prl:97:251801}
CLEO Collaboration, N.E.~Adam \etal,
\newblock Phys.\ Rev.\ Lett.{} {\bf 97},~251801~(2006)\relax
\relax
\bibitem{tech:cern-dd-ee-84-1}
R.~Brun et al.,
\newblock {\em {\sc geant3}},
\newblock Technical Report CERN-DD/EE/84-1, CERN, 1987\relax
\relax
\bibitem{pr:d57:2806}
B.W.~Harris and J.~Smith,
\newblock Phys.\ Rev.{} {\bf D~57},~2806~(1998)\relax
\relax
\bibitem{np:b374:36}
J.~Smith and W.L.~van Neerven,
\newblock Nucl.\ Phys.{} {\bf B~374},~36~(1992)\relax
\relax
\bibitem{pr:d67:012007}
ZEUS \coll, S.~Chekanov \etal,
\newblock Phys.\ Rev.{} {\bf D~67},~012007~(2003)\relax
\relax
\bibitem{pr:d27:105}
C.~Peterson \etal,
\newblock Phys.\ Rev.{} {\bf D~27},~105~(1983)\relax
\relax
\bibitem{Nason2000245}
P.~Nason and C.~Oleari,
\newblock Nucl.~Phys.{} {\bf B~565},~245~(2000)\relax
\relax
\bibitem{proc:chep:1992:222}
W.H.~Smith, K.~Tokushuku and L.W.~Wiggers,
\newblock {\em Proc.\ Computing in High-Energy Physics (CHEP), \newblock
  {Annecy, France}}, C.~Verkerk and W.~Wojcik~(eds.), p.~222.
\newblock CERN, Geneva, Switzerland (1992).
\newblock Also in preprint \mbox{DESY 92-150B}\relax
\relax
\bibitem{thesis:shehzadi:2011}
R.~Shehzadi,
\newblock Ph.D. Thesis, Universit\"at Bonn, Bonn, Germany, Report
  \mbox{BONN-IR-11-01}, 2011,
\newblock available on \texttt{http://brock.physik.uni-bonn.de/zeus\usc
  pub.php}\relax
\relax
\bibitem{epj:c1:81}
ZEUS \coll, J.~Breitweg \etal,
\newblock Eur.\ Phys.\ J.{} {\bf C~1},~81~(1998)\relax
\relax
\bibitem{thesis:briskin:1998}
G.M.~Briskin,
\newblock Ph.D.\ Thesis, Tel Aviv University, Report \mbox{DESY-THESIS
  1998-036}, 1998\relax
\relax
\bibitem{pr:d48:3160}
S.D.~Ellis and D.E.~Soper,
\newblock Phys.\ Rev.{} {\bf D~48},~3160~(1993)\relax
\relax
\bibitem{np:b406:187}
S.~Catani \etal,
\newblock Nucl.\ Phys.{} {\bf B~406},~187~(1993)\relax
\relax
\bibitem{proc:hera:1991:23}
S.~Bentvelsen, J.~Engelen and P.~Kooijman,
\newblock {\em Proc.\ Workshop on Physics at {HERA}}, W.~Buchm\"uller and
  G.~Ingelman~(eds.), Vol.~1, p.~23.
\newblock Hamburg, Germany, DESY (1992)\relax
\relax
\bibitem{proc:epfacility:1979:391}
F.~Jacquet and A.~Blondel,
\newblock {\em Proceedings of the Study for an $ep$ Facility for {Europe}},
  U.~Amaldi~(ed.), p.~391.
\newblock Hamburg, Germany (1979).
\newblock Also in preprint \mbox{DESY 79/48}\relax
\relax
\bibitem{thesis:schoenberg:2010}
V.~Sch\"onberg,
\newblock Ph.D. Thesis, Universit\"at Bonn, Bonn, Germany, Report
  \mbox{BONN-IR-10-05}, 2010,
\newblock available on \texttt{http://brock.physik.uni-bonn.de/zeus\usc
  pub.php}\relax
\relax
\bibitem{thesis:juengst:2010}
M.~J\"ungst,
\newblock Ph.D. Thesis, Universit\"at Bonn, Bonn, Germany, Report
  \mbox{BONN-IR-10-03}, 2010,
\newblock available on \texttt{http://brock.physik.uni-bonn.de/zeus\usc
  pub.php}\relax
\relax
\bibitem{thesis:yagues:2008}
A.G.~Yag\"ues Molina,
\newblock Ph.D. Thesis, Humboldt University, Berlin, Germany, Report
  \mbox{ID:6561}, 2008\relax
\relax
\bibitem{barlow:93}
R.~Barlow and C.~Beeston,
\newblock Comp.\ Phys.\ Comm.{} {\bf 77},~219~(1993)\relax
\relax
\bibitem{PhysRevD.78.013004}
P.M.~Nadolsky \etal,
\newblock Phys. Rev.{} {\bf D~78},~013004~(2008)\relax
\relax
\bibitem{springerlink:10.1140/epjc/s10052-009-1072-5}
A.D.~Martin \etal,
\newblock Eur. Phys. J.{} {\bf C~63},~189~(2009)\relax
\relax
\bibitem{Thorne:2008xf}
R.S.~Thorne, and W.K.~Tung,
\newblock Preprint \mbox{arXiv:0809.0714}, 2008\relax
\relax
\bibitem{glueck:2008}
{M.~Gl{\"u}ck}, {P.~Jimenez-Delgado} and {E.~Reya},
\newblock Eur. Phys. J.{} {\bf C~53},~355~(2008)\relax
\relax
\bibitem{Alekhin2009166}
S.~Alekhin and S.~Moch,
\newblock Phys.~Lett.{} {\bf B~672},~166~(2009)\relax
\relax
\bibitem{Alekhin:2009ni}
S.~Alekhin \etal,
\newblock Phys.\ Rev.{} {\bf D~81},~014032~(2010)\relax
\relax
\bibitem{Alekhin:2010iu}
S.~Alekhin, J.~Blumlein, S.~Moch,
\newblock PoS{} {\bf DIS2010},~021~(2010)\relax
\relax
\end{mcbibliography}
\end{document}